\documentclass[traditabstract]{aa}
\usepackage{graphicx}
\usepackage{txfonts}

\begin{document}

\title{Search for a habitable terrestrial planet transiting the nearby red dwarf GJ\,1214\thanks{The photometric time series used in this work are only available in electronic form at the CDS via anonymous ftp to  cdsarc.u-strasbg.fr (130.79.128.5) or via http://cdsweb.u-strasbg.fr/cgi-bin/qcat?J/A+A/}}

\author{M. Gillon\inst{1},  B.-O. Demory\inst{2, 3}, N. Madhusudhan\inst{4}, D. Deming\inst{5}, S. Seager\inst{2,6}, A. Zsom\inst{2}, H.~A. Knutson\inst{7},  A.~A. Lanotte\inst{1}, X. Bonfils\inst{8}, J.-M. D\'esert\inst{9}\thanks{Sagan Fellow}, L.~Delrez\inst{1}, E.~Jehin\inst{1},  J.~D. Fraine\inst{5}, P. Magain\inst{1}, A.~H.~M.~J. Triaud\inst{6}\thanks{Fellow of the Swiss National Science Foundation}}

\offprints{michael.gillon@ulg.ac.be}

\institute{
$^1$ Institut d'Astrophysique et de G\'eophysique,  Universit\'e de Li\`ege,  All\'ee du 6 Ao\^ut 17,  B\^at.  B5C, 4000 Li\`ege, Belgium \\
$^2$ Department of Earth, Atmospheric and Planetary Sciences, Department of Physics, Massachusetts Institute of Technology, 77 Massachusetts Avenue, Cambridge, MA 02139, USA\\
$^3$ Cavendish Laboratory, Department of Physics, University of Cambridge, JJ Thomson Avenue, Cambridge, CB3 0HE, UK\\
$^4$ Department of Physics and Department of Astronomy, Yale University, New Haven, CT 065, USA\\
$^5$ Department of Astronomy, University of Maryland, College Park, MD 20742-2421, USA\\
$^6$ Department of Physics and Kavli Institute for Astrophysics and Space Research, MIT, 77, Massachusetts Avenue, 
Cambridge, MA 02139, USA\\
$^7$ Division of Geological and Planetary Sciences, California Institute of Technology, Pasadena, CA 91125, USA\\
$^8$ UJF-Grenoble 1 / CNRS-INSU, Institut de Plan\'etologie et d'Astrophysique de Grenoble (IPAG) UMR 5274, Grenoble, 
F-38041, France\\
$^9$ Astronomy Department, California Institute of Technology, Pasadena, CA 91125, USA\\}

\date{Received date / accepted date}

\abstract{High-precision eclipse spectrophotometry of transiting terrestrial exoplanets represents a promising
path for the first atmospheric characterizations of habitable worlds and the search for life outside 
our solar system. The detection of terrestrial planets transiting nearby late-type M-dwarfs could make this 
approach applicable within the next decade, with soon-to-come general facilities. In this context, 
we previously identified GJ\,1214 as a high-priority target for a transit search, as the transit
probability of  a habitable planet orbiting this nearby M4.5 dwarf would be significantly enhanced by the transiting nature
of GJ\,1214\,b, the super-Earth already known to orbit the star. Based on this observation, we have set up
an ambitious high-precision photometric monitoring of GJ\,1214 with the  {\it Spitzer Space 
Telescope} to probe  the inner part of its habitable zone in search of a transiting planet as small as Mars. 
We present here the results of this transit search. Unfortunately, we did not detect any other transiting planets. Assuming
that GJ\,1214 hosts a habitable planet larger than Mars  that has an orbital period smaller than 20.9 days, 
our global analysis of the whole {\it Spitzer} dataset leads to an  a posteriori no-transit probability of $\sim 98$\%.  Our analysis 
allows us to significantly improve the characterization of GJ\,1214\,b,  to measure its occultation depth to be 70$\pm$35 ppm 
 at 4.5 $\mu$m, and to constrain it to be smaller than 205 ppm (3-$\sigma$ upper limit) at 3.6 $\mu$m. 
In agreement with the many transmission measurements published so far for GJ\,1214\,b, these  emission 
measurements  are consistent with both a metal-rich and a cloudy hydrogen-rich atmosphere.
}
\authorrunning{M. Gillon et al.}
\titlerunning{A search for a habitable planet transiting GJ\,1214}

\maketitle

\section{Introduction}

A transiting terrestrial planet orbiting in the habitable zone (HZ; Kasting et al. 1993) of 
a nearby late-type  red dwarf would represent a unique opportunity for the quest for life outside our solar system. 
It could be  suitable for the detection of atmospheric biosignatures by eclipse spectroscopy
with future facilities like the {\it James Webb Space Telescope} 
(Deming et al. 2009, Seager et al. 2009, Kaltenegger \& Traub 2009) or the European Extremely Large Telescope
(Snellen et al. 2013), thanks to a planet-to-star contrast 
and eclipse frequency that would be much more favorable than for an Earth-Sun system. 

As we outlined in a previous paper (Gillon et al. 2011a, hereafter G11), the M4.5 dwarf GJ\,1214
 represents an interesting target for attempting this detection. The MEarth ground-based transit survey 
 revealed that GJ\,1214 is transited every 1.58d by a $\sim$2.7 $R_\oplus$  super-Earth\footnote{A super-Earth is 
 loosely defined in the literature as an exoplanet of 2 to 10 Earth masses.}, GJ\,1214\,b 
 (Charbonneau et al. 2009, hereafter C09). 
 The exact nature of GJ\,1214\,b is still unknown. With a mass of $\sim$6.5  $M_\oplus$, its large radius
suggests a significant gaseous envelope that could be mainly composed
 of primordial hydrogen, making the planet a kind of mini-Neptune, or that could originate
  from the outgassing of the rocky/icy surface material of a terrestrial planet (Rogers \& Seager 2010). 
Transit transmission spectrophotometric measurements for GJ\,1214\,b  rule 
out a cloud-free atmosphere composed primarily of hydrogen, and can equally be explained by
a metal-rich composition or by a hydrogen-rich atmosphere surrounded by clouds\footnote{New HST
 data presented by Kreidberg et al. (2014) after the reviewing of this paper show unambiguously the presence of
high-altitude clouds in the atmosphere of GJ1214\,b. Still, its composition remains unknown.}
(e.g., Bean et al. 2010, 2011, D\'esert et al. 2011, Berta et al. 2012, Fraine et al. 2013, de Mooij et al. 2013)

Zsom et al. (2013) have recently presented revised values for the inner edge of the HZ of
main-sequence stars based on the extensive exploration of a large grid of atmospheric and planetary parameters. 
Based on their Eq.~12, the inner edge of the HZ of GJ\,1214 ($L\sim0.0045 L_\odot$) is $\sim$0.04 au. 
With an orbital distance of only 0.015 au, GJ\,1214\,b is not expected to be a habitable planet, but 
its only existence increases the chance that a putative second planet orbiting in the HZ 
of the host star transits it too. The members of a planetary system 
are supposed to form within a disk (e.g., Papaloizou \& Terquem 2006), so they should share nearly 
the same orbital plane, at least without any dramatic dynamical event. This assumption is  not only 
supported by the small scatter of the orbital inclinations of the eight planets in the solar 
system and of the regular satellites of its four giant planets, but also by the large 
numbers of multiple transiting systems detected by the $Kepler$ mission (Lissauer et al. 2011, 
Fabrycky et al. 2012). 

In G11, we computed that the average transit probability for a putative habitable GJ\,1214\,c was
improved by one order of magnitude thanks to the transiting configuration of GJ\,1214\,b, and we
outlined that {\it Warm Spitzer} (Stauffer et al. 2007) was the best observatory to search for this transit, 
thanks to the high photometric precision of its IRAC infrared detector (Fazio et al. 2004, Demory et al. 2011, 2012), 
and its heliocentric orbit making possible the continuous observation of most of the stars during weeks or months. 
The continuous observation of GJ\,1214 over three weeks  would probe the entire HZ of the star 
assuming an outer limit of 1.37 au for the HZ of the Sun (Kasting et al. 1993) and using an inverse-square 
law in luminosity to extrapolate the outer limit for GJ\,1214 ($0.0033 L_\odot$, C09) to be $\sim$0.08 au. 
A survey of this kind should be sensitive to planets as small as Mars for a single transit, or even smaller for multiple planets. 
This was the main concept of our {\it Warm Spitzer} program 70049 for which we present here the results of the transit search.
We also present the results of the global modeling of the entire GJ\,1214 {\it Spitzer} dataset supplemented by
ground-based data. This extensive dataset includes 21 transits and 18 occultations of GJ\,1214\,b, allowing us to 
derive strong constraints on the planet's radius and emission at  3.6 $\mu$m and 4.5 $\mu$m, and on the
 periodicity of its transits. Our detailed study of the {\it Warm Spitzer} transits of GJ\,1214\,b, and its
 implications for the transmission spectrum of the planet were presented in a separate paper (Fraine et al. 2013; 
 hereafter F13). 

 Our data and their reduction are described in Sect. 2. Our global analysis is 
presented in Sect. 3, and our search for a second planet is described in Sect. 4. We discuss our results in Sect. 5 and 
give our conclusions in Sect. 6

\section{Data description and reduction}

\subsection{{\it Spitzer} photometry}

In the context of our program 70049, {\it Spitzer} monitored GJ\,1214 continuously from 2011 April 29 03h36 UT 
to 2011 May 20 01h27 UT, corresponding to 20.9 days (502 hr) of monitoring and to the outer limit of the star's
HZ (G11). Practically, the program was divided into Astronomical Observation Requests (AORs) of 
24 hr at most,  some of them being separated by a repointing exposure. As mentioned in F13, some of the data  were
 irretrievably lost during downlink to Earth because of a Deep Space Network (DSN) ground anomaly. These 
lost  data correspond to 42 hr  of observations acquired between 12 and 14 May. The surviving data consist
 of 12383 sets of 64 individual subarray images divided in 20 AORs gathered by the IRAC detector at 4.5 
 $\mu$m with an integration time of 2 s, and calibrated by the {\it Spitzer} pipeline version S18.18.0. 
 
 In compensation for the lost observations, we were granted 42 new hours of observation that took place 
 from 2011 November 06 11h54 UT to 2011 November 08 5h47 UT. We chose to perform these new observations
 in the 3.6 $\mu$m channel, mostly to assess the dependance of the transit depth on the wavelength.  These
 data were grouped into two AORs and consist of 1166 sets of 64 individual subarray images obtained here too with 
 an integration time of 2 s, and calibrated by the {\it Spitzer} pipeline version S19.1.0. Our {\it Spitzer}
 data are available on the {\it Spitzer} Heritage Archive database\footnote{http://sha.ipac.caltech.edu/applications/Spitzer/SHA}.
 
 We complemented our data set with all the other {\it Spitzer} data publicly available on the {\it Spitzer} Heritage 
 Archive database for GJ\,1214\,b, including two transits observed respectively at 3.6 $\mu$m and 4.5 $\mu$m in  
 the program 542 (PI D\'esert, D\'esert et al. 2011), and six occultations (three at 3.6 $\mu$m and three at 4.5 $\mu$m)  observed in 
 the program 70148 (PI Madhusudhan). The logs of our photometric data set are given in Table 1.
 
We used the following reduction strategy  for all the {\it Spitzer} data. We first converted fluxes from the 
{\it Spitzer} units of specific intensity (MJy/sr) to photon counts, and then  we performed aperture photometry  on 
each subarray image with the {\tt IRAF/DAOPHOT}\footnote{IRAF is distributed by the National Optical Astronomy 
Observatory, which is operated by the Association of Universities for Research in Astronomy, Inc., under cooperative 
agreement with the National Science Foundation.} software (Stetson, 1987). We tested different aperture radii and 
background annuli, obtaining better results  with an aperture radius of 2.5 pixels and a background annulus 
extending from 11 to 15.5 pixels  from the  point-spread function (PSF) center. For the first two 3.6 $\mu$m AORs taken 
in program 70148, we obtained a better result with an aperture of 2.75 pixels. We measured the center and width of the PSF by 
fitting a 2D-Gaussian profile on each image. We then looked at the $x$-$y$ distribution of the measurements,
 and we discarded the few measurements having a visually discrepant position relative to the bulk of the data. For each block 
of 64 subarray images, we then discarded the discrepant values for the measurements of flux, background, $x$ and 
$y$ positions, and  PSF widths in the $x$- and $y$-direction, using a 10-$\sigma$ median clipping  
for the six parameters. We averaged the remaining values, taking the errors on the average
 flux measurements as photometric errors. At this stage, we used a moving median filter in flux on the resulting light curve to
 discard outlier measurements due to cosmic hits, for example.  Finally,
we discarded from the second 3.6 $\mu$m AOR of our program 70049  two blocks of $\sim$1 hr duration 
corresponding to sharp flux  increases of $\sim$500 ppm followed by  smooth decreases  to the normal level. We attribute
these structures to the effect of cosmic hits on the detector. In the end, $\sim$ 5\% and 0.5\% of the measurements were discarded
 at 3.6 $\mu$m and 4.5 $\mu$m, respectively.

\subsection{TRAPPIST transit photometry}

In 2011, we observed seven transits of GJ\,1214\,b from Chile with the 60 cm robotic telescope 
TRAPPIST\footnote{see http://www.ati.ulg.ac.be/TRAPPIST} ($TRA$nsiting $P$lanets
and $P$lanetes$I$mals $S$mall $T$elescope; Gillon et al. 2011b, Jehin et al. 2011) located
at ESO La Silla Observatory. TRAPPIST is equipped with a thermoelectrically-cooled 2k $\times$ 2k CCD 
camera. Its field of view is 22' $\times$ 22'.
We monitored all the transits with the telescope slightly defocused
and in the $I+z$ filter that has a transmittance $>90$ \% from 750 nm to beyond 1100 nm. We used an 
exposure time of 25 s for all integrations, the read-out + overhead time being $\sim$5 s.
 Three of the transits observed  by {\it Spitzer} were also observed by TRAPPIST. 

After a standard pre-reduction (bias, dark, flat-field correction), we extracted the stellar fluxes 
from the TRAPPIST images using {\tt IRAF/DAOPHOT}. We tested several sets of reduction parameters, and 
we kept the one giving the most precise photometry for the star of similar brightness to GJ\,1214. 
After a careful selection of 13 reference stars, differential photometry was then obtained. Table 2
provides the logs of these TRAPPIST data.

\subsection{TRAPPIST variability photometry}

In addition to the seven transits mentioned above, TRAPPIST monitored  GJ\,1214 regularly
from 2011 April 19 to 2011 May 19. These observations consisted of blocks of a few exposures taken
in the $Ic$ filter, their goal being to assess the global variability of the star during the {\it Spitzer}
 survey. The resulting photometry was not used as input data in our global analysis described in the next section.
  The reduction procedure was similar to the one used for the transits. Six
comparison stars were carefully selected on the basis of their stability during the covered month. 
For each comparison star, we determined a red noise value on a timescale of 24 hr by following 
the procedure described in Gillon et al. (2006), as was done by Berta et al. (2011) in their 
study of the variability of GJ\,1214. Averaging the values of all the comparison stars, we obtained
a mean red noise value of 0.1 \% that we added quadratically to the errors on the average flux 
measured on GJ\,1214 for each night. The resulting light curve for GJ\,1214 is visible in Fig. 1. It 
shows no obvious flux variation, its $rms$ being $\sim$0.15 \%, equal to the mean error. 
We conclude from this light curve that the star was quiet at the 1-2 mmag level in the $Ic$ filter 
during the $\it Spitzer$ run. As the photometric variability of GJ\,1214 is driven by spots rotating with
its surface (Berta et al. 2011), its amplitude must decrease with increasing wavelength. Assuming
spots 300-500 K cooler than the mean photosphere leads to the conclusion that the star was 
stable at the 0.5-1 mmag level in the 4.5 $\mu$m channel during our main {\it Spitzer} run of 
three weeks.

\begin{figure}
\label{fig:1}
\centering                     
\includegraphics[width=9cm]{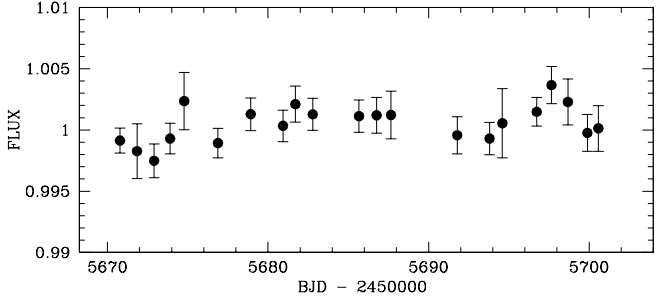}
\caption{Variability light curve data gathered between 2011 April 19 and 2011 May 19 in the $Ic$ filter with 
the TRAPPIST telescope. }
\end{figure}

\begin{table*}
\begin{center}
\label{tab:gj1214_1}
\begin{tabular}{ccccccccccc}
\hline\noalign {\smallskip}
Program & AOR         & Start & IRAC band  & $N_p$ & Baseline  &  BLISS   &  BLISS & $\beta_w$ & $\beta_r$   \\ \noalign {\smallskip} 
 ID &  ID  &  date & ($\mu$m)    &              &   model      &  $N_x$  &  $N_y$& &    \\ \noalign {\smallskip}
\hline \noalign {\smallskip}
542         & 39218176  &  2010 Apr 26    & 3.6 & 101 &  $p([xy]^3+w_x^1)$              & 0 & 0  & 0.92 & 1.06  \\ \noalign {\smallskip}
542         & 39217920  &  2010 Apr 27    & 4.5 & 103  &  $p([xy]^2+l^2)$                   & 0 & 0  & 0.85 & 1.20 \\ \noalign {\smallskip}
70148     & 40216832  & 2010 Oct 16    & 3.6  & 110 & $p([xy]^2+w_y^3+l^2)$       & 0 & 0  & 0.93 & 1.56 \\ \noalign {\smallskip}
70148     & 40217088  & 2010 Oct 17    & 4.5 & 109 & $p([xy]^2+w_x^1)$                & 0 & 0  & 0.85  & 1.33\\ \noalign {\smallskip}
70148     & 40217344  & 2010 Oct 25    & 3.6  & 109 & $p([xy]^2+w_x^1+w_y^1)$ & 0 & 0 & 0.66  & 1.15  \\ \noalign {\smallskip}
70148     & 40217600  & 2010 Oct 27    & 4.5  & 109 & $p([xy]^2+w_x^1)$               & 0 & 0  & 1.04 & 1.05 \\ \noalign {\smallskip}
70148     & 40217856 & 2010 Oct 28     & 3.6 & 110 & $p([xy]^2+l^2)$                      & 0 & 0  & 0.93  & 1.00 \\ \noalign {\smallskip}
70148     & 40211882 & 2010 Oct 31     & 4.5 & 109 & $p([xy]^2+w_x^2)$                & 0 & 0  & 0.96 & 1.59 \\ \noalign {\smallskip}
70049   & 42045952   & 2011 Apr 29    & 4.5 & 666  & $p([xy]^2+w_x^1+w_y^2)$ & 10 & 11 & 0.93  & 1.17 \\ \noalign {\smallskip}                            
70049   & 42046208   & 2011 Apr 30    & 4.5 & 667  & $p([xy]^2+w_y^2)$               &  11  & 10  & 0.99 & 1.20\\ \noalign {\smallskip}                            
70049   & 42046464   & 2011 May 1    & 4.5 & 597  & $p([xy]^2+w_x^1)$                & 9 & 9 & 0.98 &  1.37 \\ \noalign {\smallskip}                           
70049   & 42046720   & 2011 May 2    & 4.5 & 663  & $p([xy]^2+w_y^2)$                & 9 & 10  & 0.99  & 1.15 \\ \noalign {\smallskip}                           
70049   & 42046976   & 2011 May 3    & 4.5 & 667  & $p([xy]^2+w_x^1+w_y^2)$  & 10 & 9  & 0.96 & 1.40\\ \noalign {\smallskip}                                                                
70049   & 42047232   & 2011 May 4    & 4.5 & 610  & $p([xy]^2+w_x^2)$               & 10 & 9  & 0.94  & 1.19 \\ \noalign {\smallskip}                                                                
 70049   & 42047488   & 2011 May 5    & 4.5 & 665  & $p([xy]^2+w_x^2+w_y^1)$  & 9 & 10 & 0.92 & 1.17 \\ \noalign {\smallskip}                                                                
70049   & 42047744   & 2011 May 6    & 4.5 & 667  & $p([xy]^2+w_x^2+w_y^1)$  & 9 & 9 & 0.87  & 1.07\\ \noalign {\smallskip}                                                                
70049   & 42048000   & 2011 May 7    & 4.5 & 472  & $p([xy]^2+w_y^1)$                & 8 & 8  & 0.93 & 1.74 \\ \noalign {\smallskip}                                                                
70049   & 42048256   & 2011 May 8    & 4.5 & 667  & $p([xy]^2)$                              & 10 & 9 & 0.94 &  1.03 \\ \noalign {\smallskip}                                                                
70049   & 42048512   & 2011 May 9    & 4.5 & 666  & $p([xy]^2+w_x^1)$                & 10 & 9 & 0.98 & 1.21 \\ \noalign {\smallskip}                                                                
70049   & 42048768   & 2011 May 10   & 4.5 & 667  & $p([xy]^2+w_x^1)$               & 10 & 10 & 0.98 & 1.81 \\ \noalign {\smallskip}                                                                
70049   & 42049280   & 2011 May 11    & 4.5 & 666  & $p([xy]^2+w_x^1+w_y^1)$  & 9 & 9 & 0.89 & 1.18 \\ \noalign {\smallskip}                                                                
70049   & 42049536   & 2010 May 12    & 4.5 & 112  & $p([xy]^2+w_x^1+w_y^1)$  & 0  & 0 & 0.91  & 1.16 \\ \noalign {\smallskip}                                                                
70049   & 42050048   & 2011 May 14    & 4.5 & 655  & $p([xy]^2+w_y^3)$                & 10 & 9 & 0.98 & 1.77 \\ \noalign {\smallskip}                                                                
70049   & 42050304   & 2011 May 15    & 4.5 & 667  & $p([xy]^2+w_x^1+w_y^1)$  & 10 & 10 & 0.95 & 1.16 \\ \noalign {\smallskip}                                                                
70049   & 42050560   & 2011 May 16   & 4.5 & 583  & $p([xy]^2+w_x^1+w_y^1)$  & 10 & 9 & 0.83 & 1.27 \\ \noalign {\smallskip}                                                                
70049   & 42050816   & 2011 May 17    & 4.5 & 667  & $p([xy]^2+w_y^2)$                & 11 & 10 & 0.90 & 1.45 \\ \noalign {\smallskip}                                                                
70049   & 42051072   & 2011 May 18    & 4.5 & 667  & $p([xy]^2+w_y^2)$                & 10 & 10 & 0.89 & 1.12 \\ \noalign {\smallskip}                                                                
70049   & 42051328   & 2011 May 19    & 4.5 & 639  & $p([xy]^2+w_x^1)$                & 10 & 9 & 0.89 & 1.67 \\ \noalign {\smallskip}                                                                
70049     & 44591872  & 2011 Nov 6     & 3.6 & 660  & $p([xy]^2+w_x^1+w_y^3+l^1)$ & 11 & 10 & 0.89 &  1.54 \\ \noalign {\smallskip}
70049     & 44592128  & 2011 Nov 7     & 3.6 & 443  & $p([xy]^2+w_x^1+w_y^3)$        & 7 & 8  & 0.82 & 1.29 \\ \noalign {\smallskip}
\hline 
\end{tabular}
\caption{{\it Spitzer} light curves used in this work. Each light curve corresponds to a specific {\it Spitzer} observing block (AOR). For each of them, the table gives the ID of the {\it Spitzer} program and of  the AOR, the start date, the IRAC channel used, the number of measurements, the baseline function selected for our global
modeling (see Sect. 3), the number of divisions in the $x$- and $y$-directions used for the BLISS pixel mapping (see Sect. 3), and the
 $\beta_w$ and $\beta_r$ 
error rescaling factors (see Sect. 3). For the baseline function, $p(\epsilon^N)$ denotes, respectively, a $N$-order polynomial function of the logarithm of time 
($\epsilon=l$), of the PSF $x$- and $y$-positions ($\epsilon=[xy]$), and widths ($\epsilon=w_x$  \& $w_y$). }
\end{center}
\end{table*}

\begin{table}
\begin{center}
\label{tab:gj1214_2}
\begin{tabular}{cccccc}
\hline\noalign {\smallskip}
 Date  & Filter & $N_p$ & Baseline  & $\beta_w$ & $\beta_r$ \\ \noalign {\smallskip} 
            &           &              &   model    &   &  \\ \noalign {\smallskip}
\hline \noalign {\smallskip}
2011 Mar 11 & $I+z$ & 195 & $p(t^2)$   & 0.76 &  1.19     \\ \noalign {\smallskip}
2011 Mar 30 & $I+z$&  248 & $p(t^2)$   & 0.83 &   1.79    \\ \noalign {\smallskip}
2011 Apr 18 & $I+z$ & 234 & $p(t^2)$   & 0.71  &   1.02    \\ \noalign {\smallskip}
2011 Apr 26 & $I+z$ & 169 & $p(t^2)$   & 0.93  &   1.09    \\ \noalign {\smallskip}
2011 Apr 29 & $I+z$ & 286 & $p(t^2)$    & 0.99  &   1.54   \\ \noalign {\smallskip}
2011 May 15 & $I+z$ &  224 & $p(t^2)$ & 0.84   &   1.29    \\ \noalign {\smallskip}
2011 May 18 & $I+z$ & 303 & $p(t^2)$  & 0.85   &    1.00  \\ \noalign {\smallskip}
\hline 
\end{tabular}
\caption{TRAPPIST  transit light curves used in this work. For each light curve, this table gives the  date, 
the filter used, the number of measurements, the baseline function used in our global modeling, and the $\beta_w$ and $\beta_r$ 
error rescaling factors (see Sect. 3). For the baseline function, 
$p(t^2)$ denotes a second-order polynomial function of time. }
\end{center}
\end{table}
 
\section{Global data analysis}

We performed a global analysis of our extensive photometric dataset, and used the resulting residuals of the best-fit model 
as input data for our transit search (Sect. 4). We describe here the global analysis.

\subsection{Method and model}

Our data analysis was based on the most recent version of our adaptive Markov Chain
Monte-Carlo (MCMC) algorithm described in detail in Gillon et al. (2012). The assumed
model consisted in using the eclipse model of Mandel \& Agol (2002) to represent the
transits and occultations of GJ\,1214\,b, multiplied by  a phase curve model for 
both  {\it Spitzer} channels, and 
multiplied for each light curve by a baseline model aiming to represent the other astrophysical 
and instrumental mechanisms able to produce photometric variations. We assumed a quadratic  limb-darkening 
law for the transits. For each light curve corresponding to a specific AOR, we based the selection of the baseline model
 on the minimization of the Bayesian information criterion (BIC; Schwarz 1974) 
as described in Gillon et al. (2012). Tables 1 and 2 present the baseline function elected for each light curve. 

For the {\it Spitzer} photometry, our baseline models included three types of low-order polynomials: \begin{itemize}
\item one representing the dependance of the fluxes to the $x$- and $y$-positions of the  PSF center.
 This model represents the well-documented pixel phase effect 
on the IRAC InSb arrays (e.g., Knutson et al. 2008). 
\item one representing a dependance of the fluxes to the   PSF widths in the $x$- and/or $y$-direction. 
Modeling this dependance was required for most light curves. Considering the under-sampling of the  PSF
(full-width at half maximum $\sim$ 1.5 pixels) and the significant inhomogeneity of the response within each pixel, 
variations of the measured  PSF width  correlated with the wobble of its center are to be expected. 
Still, the need for a model relating the fluxes and the  PSF widths suggests an actual variability of  the 
 PSF, otherwise its measured width and position should be totally correlated
 and the effects of the  PSF width variations would be corrected by the pixel phase model. 
\item one representing a sharp increase of the detector response at the start of some AORs and modeled
with a polynomial of the logarithm of time. This model was required 
only for four AORs, three taken at 3.6 $\mu$m and one at 4.5 $\mu$m. This ramp effect is also well-documented (e.g., Knutson et al. 2008)
and is attributed to a charge-trapping mechanism resulting in a dependance of the pixels' gain on their illumination history. The
effect was much stronger for the SiAs IRAC arrays (5.8 $\mu$m and 8 $\mu$m); it also affects the InSb arrays, but to a lesser 
extent.
\end{itemize}

For the shorter AORs taken in programs 542 and 70148, the pixel phase effect was well represented by 
a low-order polynomial of the $x$ and $y$ PSF center positions. For the AORs taken in our program 70049 with a 
typical duration of 24~h, the excursions of the  PSF center were larger and better results were obtained by complementing the 
position polynomial model with the Bi-Linearly-Interpolated Sub-pixel Sensitivity (BLISS) mapping method presented by 
Stevenson et al. (2012). This method uses the data themselves to map the intra-pixel sensitivity at high resolution at each step
of the MCMC. In our 
implementation of the method, the detector area probed by the  PSF center for a given AOR is divided into $N_x$ and $N_y$ 
slices along the $x$- and $y$-directions, respectively. The values $N_x$ and $N_y$ are selected so that ten measurements
on average fall within the same sub-pixel box.  This last criterion was chosen empirically 
to model properly the higher frequencies of the sensitivity map while avoiding overfitting the data with too few 
measurements per sub-pixel box (i.e., too many degrees of freedom). All the other aspects of our 
implementation of the method are similar to the ones presented by Stevenson et al. (2012) and we refer the reader to their 
paper for more details. Table 1 gives the number of divisions in the $x$- and $y$-directions used for the BLISS-mapping 
for each {\it Spitzer} light curve. 

It can be noticed from Table 1 that our baseline models represent only {\it Spitzer} systematic effects, and do not contain 
any term representing a possible stellar variability (e.g., a linear trend). For each light curve, we systematically tested more complex 
baseline models with time dependance, but the resulting model marginal likelihoods as estimated from the BIC
were poorer in all cases. This indicates a very low level of variability for GJ\,1214, in excellent agreement with 
 our $Ic$ light curve obtained with TRAPPIST (Sect. 2.3, Fig.~1). 

For each {\it Spitzer} channel,  the assumed phase curve model was the sinus function \begin{equation}
F_{phase, i} = 1 - A_i \cos \bigg( \frac{2\pi(t-T_0)}{P} - O_i \bigg),
\end{equation} where $i$ is 3.6  $\mu$m or 4.5 $\mu$m,  $t$ is the time, $T_0$ and $P$ are the time of inferior conjunction and the
orbital period of GJ\,1214\,b, and the parameters $A_i$ and $O_i$ are the semi-amplitude and 
phase offset of the phase curve. As no phase effect could be detected with this simple function (see below), we did not 
test more sophisticated phase curve models.

After election of the baseline model for each light curve, 
we performed a preliminary global MCMC analysis of our extensive data set, following the procedure 
described in Gillon et al. (2012). A circular orbit was assumed for GJ\,1214\,b. 
The parameters that were randomly perturbed at each step of the Markov 
chains (called {\it jump parameters}) were \begin{itemize}
\item the stellar mass $M_\ast$, assuming a normal prior distribution corresponding to  $0.176 \pm 0.009 M_\odot$, 
the value and error recently presented  by Anglada-Escud\'e et al. (2013, hereafter AE13)  from infrared apparent magnitudes 
and their updated parallax of the star combined with  empirical relations between $JHK$ absolute magnitudes and 
stellar mass (Delfosse et al. 2000); 
\item the stellar effective temperature $T_{eff}$ and metallicity [Fe/H], assuming the normal prior distributions  
corresponding to $T_{eff} = 3250\pm 100$ K and [Fe/H] = $0.1\pm0.1$ based on the results of AE13; 
\item the planet/star area ratio $dF_i =(R_p /R_\star )^2$ for the three probed channels ($I+z$, 3.6  $\mu$m and 4.5 $\mu$m), 
$R_p$ and $R_\star$ being, respectively, the radius of the planet and the star; 
\item the occultation depths $dF_{occ, i}$ at 3.6  $\mu$m and 4.5 $\mu$m;
\item the parameter $b' = a \cos{i_p}/R_\star$, which is the transit impact parameter in the case of a 
circular orbit, $a$ and $i_p$ being, respectively, the semi-major axis and inclination of the orbit;
\item the orbital period $P$; 
\item the time of inferior conjunction $T_0$;
\item the transit width (from first to last contact) $W$; 
\item the phase curve parameters $A_i$ and $O_i$ for the 3.6  $\mu$m and 4.5 $\mu$m {\it Spitzer} channels. 
For each channel, the phase curve semi-amplitude was forced to be equal to or smaller than 
half of the occultation depth $dF_{occ}$.
\end{itemize} For each bandpass, the two quadratic limb-darkening coefficients $u_1$ and $u_2$ were also let free, 
using as jump parameters not these coefficients themselves but the combinations $c_1 = 2  \times u_1  + u_2$  and 
$c_2 = u_1 - 2 \times u_2$ to minimize the correlation of the obtained uncertainties. Normal prior distributions
were assumed for $u_1$ and $u_2$;  the corresponding expectations and standard deviations were
interpolated from the tables of Claret \& Bloemen (2011) for the corresponding bandpasses and for
 $T_{eff} = 3250\pm 100$ K, $\log{g} = 5.0 \pm 0.1$, and [Fe/H] = $0.1\pm0.1$  (AE13). 
 
This preliminary global MCMC analysis allowed us to
assess the need for rescaling the photometric errors. The $rms$ of the residuals was compared to the
mean photometric errors, and the resulting factor $\beta_w$ were stored;  $\beta_w$ represents the
under- or overestimation of the white noise of each measurement. The red noise present in the
light curve (i.e., the inability of our model to represent  the data perfectly) was taken into account as 
described by Gillon et al. (2010), in other words, a scaling factor $\beta_r$ was determined from the $rms$ of the 
binned and unbinned residuals for different binning intervals ranging from 5 to 90 minutes, the largest values
being kept as $\beta_r$. In the end, the error bars were multiplied by the correction factor $CF = \beta_r 
\times \beta_w$. The values of $\beta_w$ and $\beta_r$ derived  for each light curve are given in 
Tables 1 and 2. 

One can notice that most $\beta_w$ are smaller than 1. For {\it Spitzer}, each of our measurements is
 the mean of 64 individual measurements, and
 our selected photometric errors are the errors on this mean. It is normal that this procedure slightly overestimates
 the actual photometric error, as the wobbles of the telescope pointing have frequencies high enough
to lead to significant  PSF position variations during a block of 64 measurements ($64 \times 2$ s = 128 s), increasing
the scatter of the individual measurements because of the phase pixel effect. For TRAPPIST, the $\beta_w$
smaller than 1 are probably due to an overestimation of the scintillation noise, derived in TRAPPIST pipeline 
from the usually quoted formula of Young (1967). One can also notice that our derived  $\beta_r$ are relatively close
to 1 (mean values of 1.27, 1.31, and 1.27 for {\it Spitzer} at 3.6 $\mu$m and 4.5 $\mu$m, and for TRAPPIST, respectively), 
revealing low levels of red noise in our residual light curves.

\subsection{Analysis assuming a circular orbit}

 In a first step, a circular orbit was assumed for GJ\,1214\,b, based on  the recent analysis of 
a set of 61 radial velocities gathered with the HARPS spectrograph (X. Bonfils, in prep.) that resulted in 
an orbital solution fully consistent with a circular orbit, the eccentricity $e$ being constrained to be
smaller than 0.12 with 95\% confidence. 

Our MCMC analysis consisted of two chains of 100,000 steps. 
Its convergence was successfully checked through the statistical test of Gelman \& Rubin (1992). Its main results are 
shown in Table 3 (MCMC 1) that gives the deduced values and error bars for the jump and system parameters. 
Figure 2 shows the photometry acquired in our program 70049 with the best-fit global models superimposed. 
It also shows the photometry corrected for {\it Spitzer} systematic effects.
 Figure 3 shows the best-fit transit and 
occultation models superimposed on the period-folded photometry  for the three channels probed by our data, after division 
by the best-fit baseline + phase curve model.
Figure 4 shows the folded and detrended {\it Spitzer} photometry with the best-fit eclipses + phase curve models.

Several conclusions can be drawn from the results shown in Table 3.\begin{itemize}
\item The transit depths deduced for the three channels are consistent with each other.
\item For both {\it Spitzer} channels, the phase effect is not detected and we can only put upper limits
on its amplitude.
\item The occultation of the planet is not detected at 3.6 $\mu$m, and its amplitude is constrained to be $<$205 ppm 
(3-$\sigma$ upper limit). Assuming for the star spectral energy distribution a spectrum model of Kurucz (1993) with
local thermodynamic equilibrium, $T_{eff} = 3170$ K, 
$[Fe/H] = 0$, and $\log{g} = 5.0$, we derive from this upper limit a maximum brightness temperature of 850 K.
At 4.5 $\mu$m, the occultation is detected at the 2-$\sigma$ level, its derived depth value of $70\pm35$ ppm corresponding
to a brightness temperature of $545_{-55}^{+40}$ K.
 \end{itemize} 
These results are discussed more thoroughly in Sect. 5.

\begin{figure*}
\label{fig:2}
\centering                     
\includegraphics[width=8.5cm]{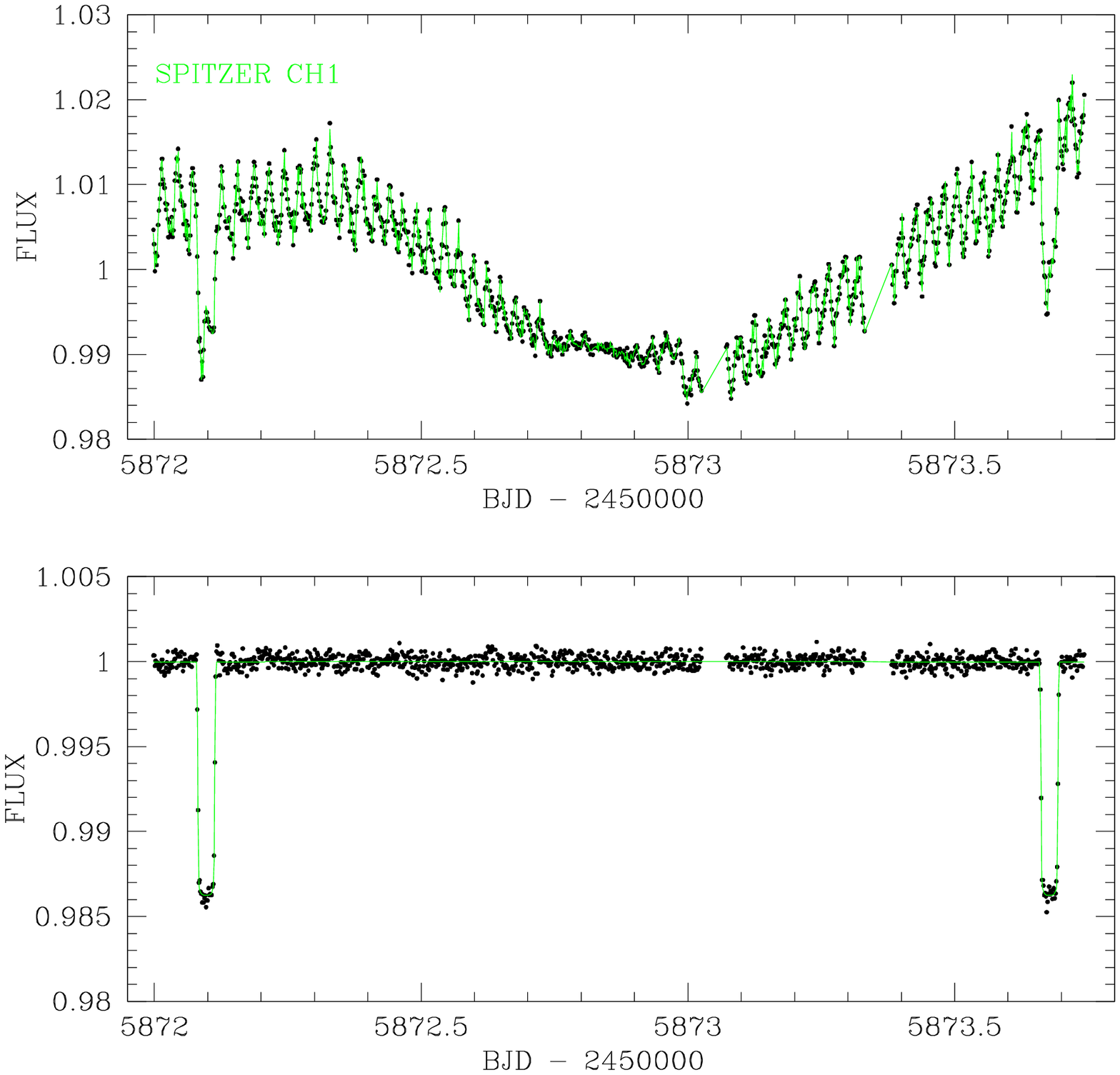}
\includegraphics[width=8.5cm]{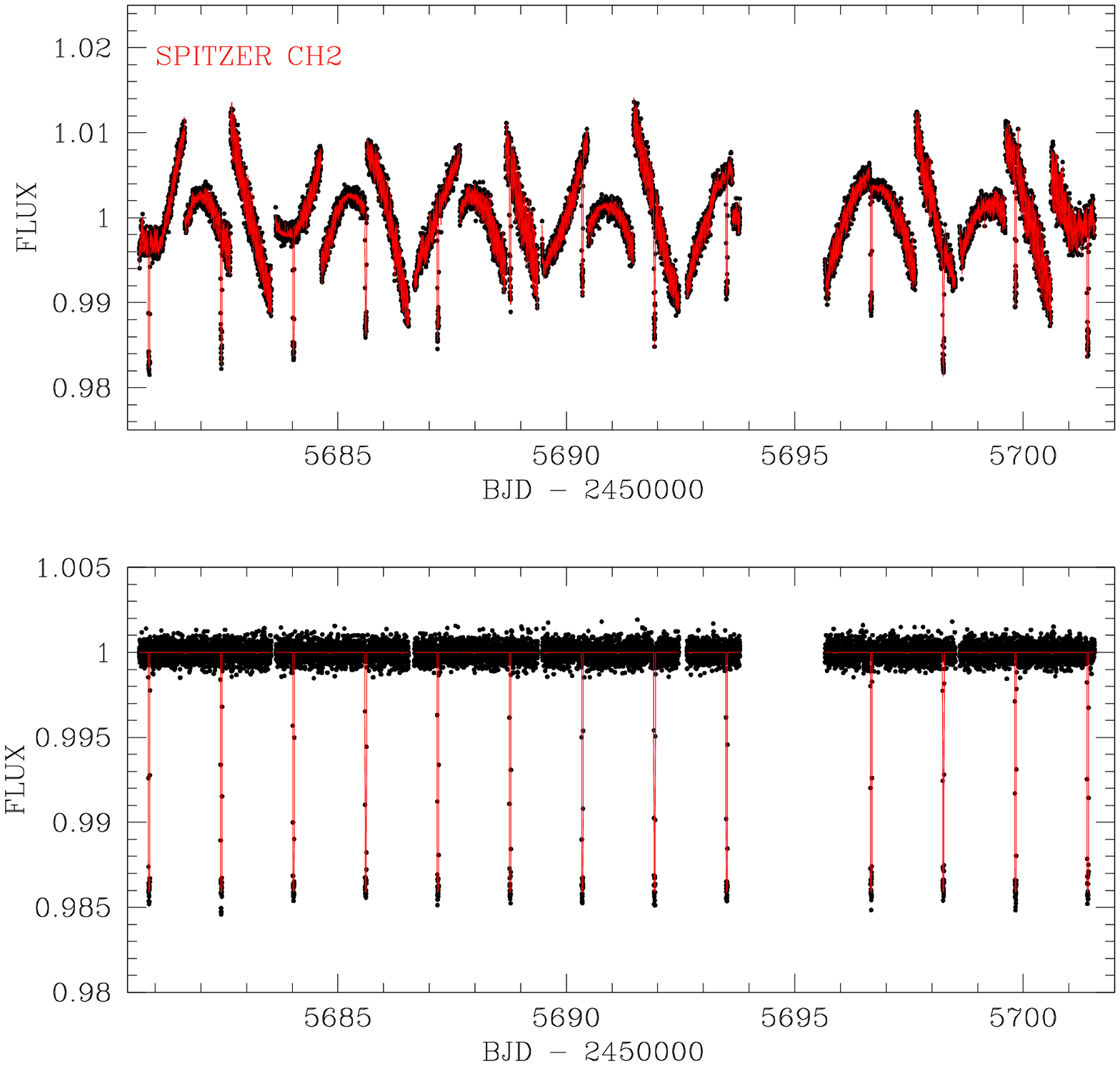}
\caption{$Top$: 3.6 $\mu$m  ($left$) and 4.5 $\mu$m  ($right$) light curves obtained in our program 70049 for GJ\,1214, 
with the best-fit global models (Sect. 3.2) for each AOR superimposed.  $Bottom$: The same after division by the best-fit instrumental models.}
\end{figure*}

\begin{figure*}
\label{fig:3}
\centering                     
\includegraphics[width=9cm]{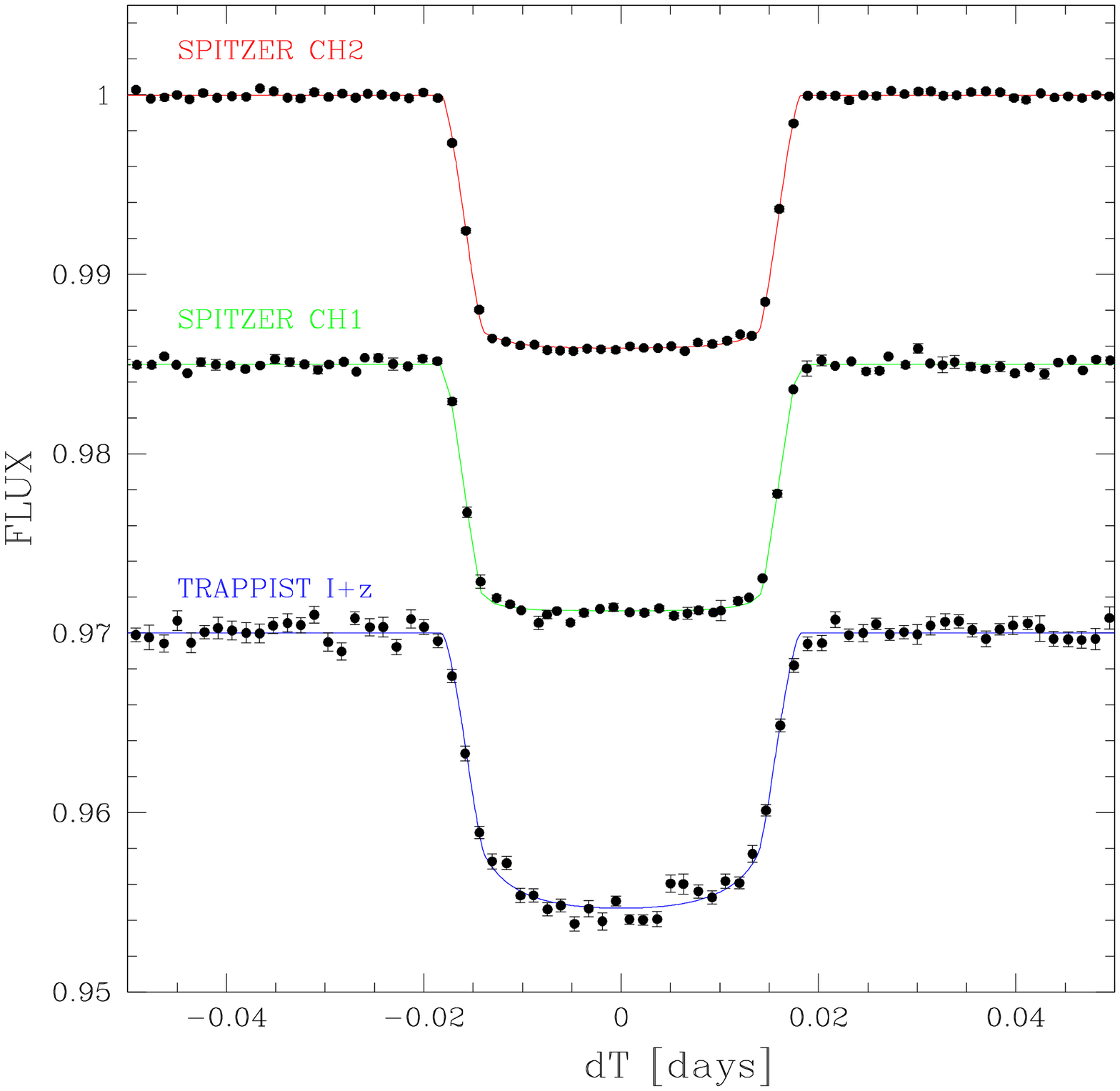}
\includegraphics[width=9cm]{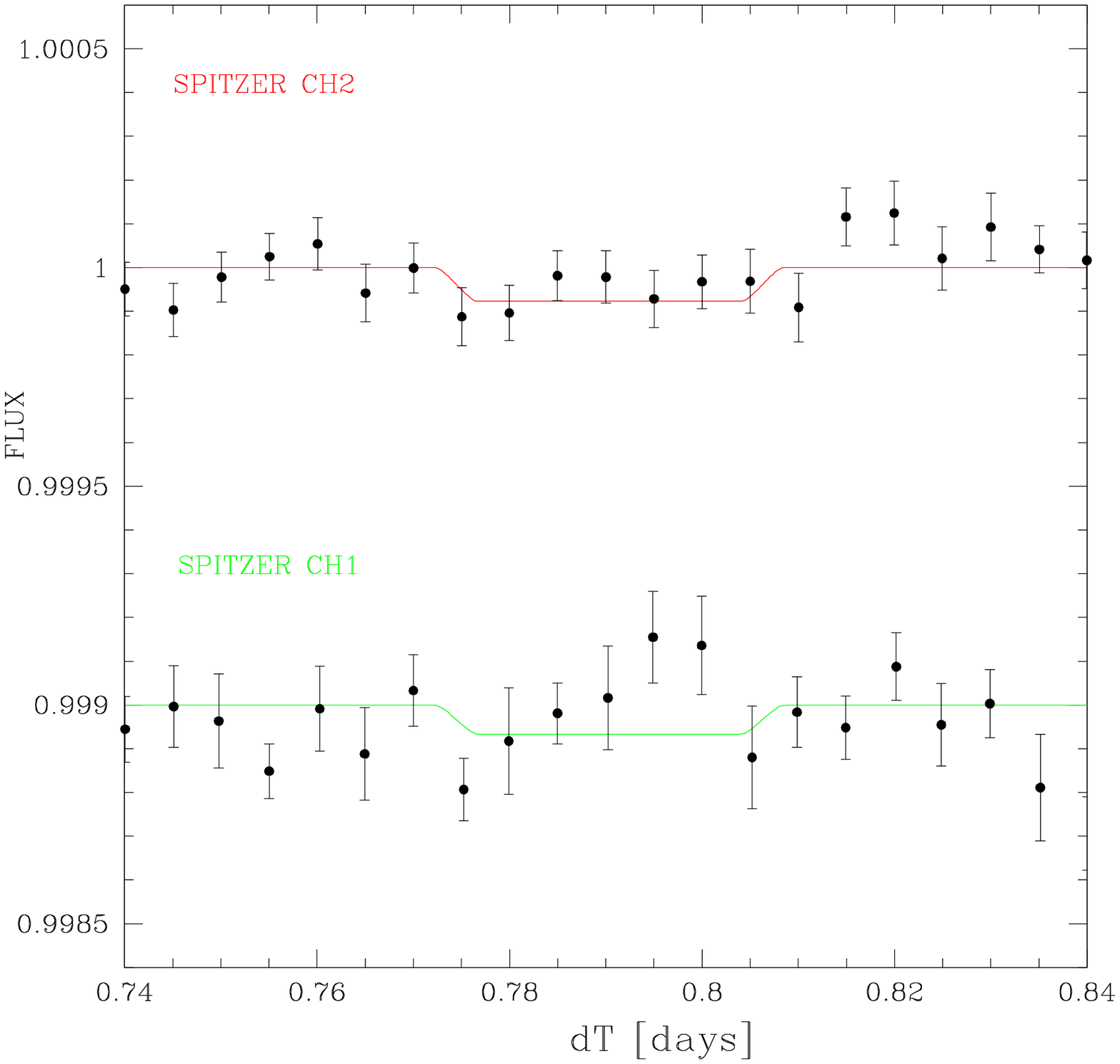}
\caption{Detrended photometry period-folded on the best-fit transit ephemeris obtained in our 
global analysis (Sect. 3.2)  after zoom on the transit ($left$) and occultation ($right$) phases. 
For the transit and occultation phases, the measurements were binned per interval of 2 min and 7.2 min, 
respectively. For both panels, the best-fit eclipse models are superimposed. }
\end{figure*}

\begin{figure}
\label{fig:4}
\centering                     
\includegraphics[width=9cm]{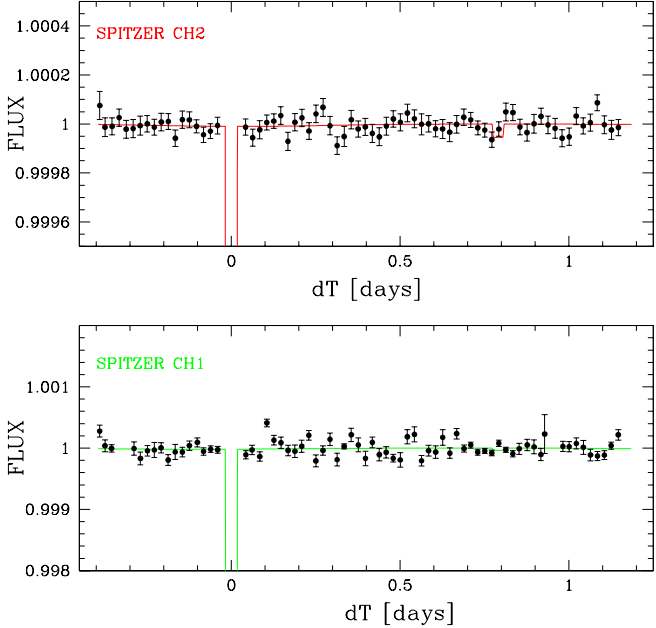}
\caption{Detrended {\it Spitzer} 3.6 $\mu$m ($top$) and 4.5 $\mu$m ($bottom$) photometry folded on the best-fit transit ephemeris
obtained in our global analysis (Sect. 3.2), binned per intervals of 30 min. The best-fit eclipse + phase-curve models are superimposed.}
\end{figure}

\subsection{Analysis assuming an eccentric orbit}

As outlined by Carter et al. (2011), the circularization timescale of GJ\,1214\,b could be as long as 10 Gyr for 
specific compositions, while several transiting Neptune-like planets have significantly eccentric orbits. Even 
when considering the new HARPS measurements, a small orbital eccentricity is still possible, so based on these
considerations it is desirable to assess the influence of the circular orbit assumption on the stellar and planetary
size, and on the planet's thermal emission. To carry out this task, we performed a second MCMC analysis with the
 orbital eccentricity $e$ and argument  of pericenter $\omega$ free, the corresponding jump parameters being 
 $\sqrt{e}\cos\omega$ and $\sqrt{e}\sin\omega$. Gaussian prior probability distributions were assumed for these
  two jump parameters, based on the values $\sqrt{e}\cos\omega = 0 \pm 0.12$ and $\sqrt{e}\sin\omega = -0.10
   \pm 0.17$ deduced from the analysis of the new HARPS dataset. The corresponding distributions for $e\cos\omega$ 
   and $e\sin\omega$ are, respectively, $0.00 \pm 0.02$ and $-0.02_{-0.06}^{+0.03}$. 

The results of this second analysis are given in Table 3 (MCMC 2). It can be seen that the derived parameters for the
 system are in good agreement with the ones deduced under the circular orbit assumption, but some are 
 less precise because the uncertainties on $e$ and $\omega$ propagate to the parameters $a/R_\ast$, 
 $\rho_\ast$, $R_\ast$, and $R_p$. We note, however, that our adopted results are the ones from the analysis  assuming a circular orbit, based
 on the absence of observational evidence for a significant eccentricity ($\Delta$BIC = -20 in favor of the circular model).
 
\subsection{Analysis with a uniform prior distribution on the limb-darkening coefficients}

We explored the influence of our selected priors on the limb-darkening  by performing
a MCMC analysis assuming for all channels uniform prior distributions on the quadratic limb-darkening coefficients $u_1$ and $u_2$.
Its results are presented in Table 3 (MCMC 3). While some derived values are slightly less precise than the ones obtained in our adopted analysis, 
they are in excellent agreement ($< 1 \sigma$) with them, including for the parameters defining the transit shape ($dF$, $b'$, $W$, $\rho_\ast$, 
$a/R_\ast$, $i$). We thus conclude that the results of our adopted analysis are not influenced by our selected priors 
on the limb-darkening coefficients. 

\subsection{Analysis allowing for transit timing variations}

In a final MCMC analysis, we let the timings of the transits present in our {\it Spitzer} + TRAPPIST dataset be  jump parameters.
Our goal was to benefit from the strong constraints brought by the global analysis on the transit shape and depth to 
reach the highest possible sensitivity on possible transit timing variations (TTVs, Holman \& Murray 2005, Agol et al. 2005) due to another unknown object in the system. 
In this analysis, we assumed a circular orbit for 
GJ\,1214\,b, and normal prior distributions based on the results of our adopted analysis  (Table 3, MCMC 1) for the jump parameters $P$ and $T_0$.

Table 4 presents the derived transit timings. A linear regression using these timings and their epochs as input led to the following 
transit ephemeris: $2454980.748996 (\pm0.000084) + N \times 1.58040418 (\pm 0.00000019)$ BJD$_{TDB}$, $N$ being the epoch.
This ephemeris agrees well with the MCMC result (Table 3). 

Figure 5 shows the resulting TTVs as a function of the epochs of the transits. As can be seen 
in this figure, we could not detect any significant TTV, which is consistent with the results 
that we independently obtained in F13 from the same data.

\begin{figure}
\label{fig:5}
\centering                     
\includegraphics[width=9cm]{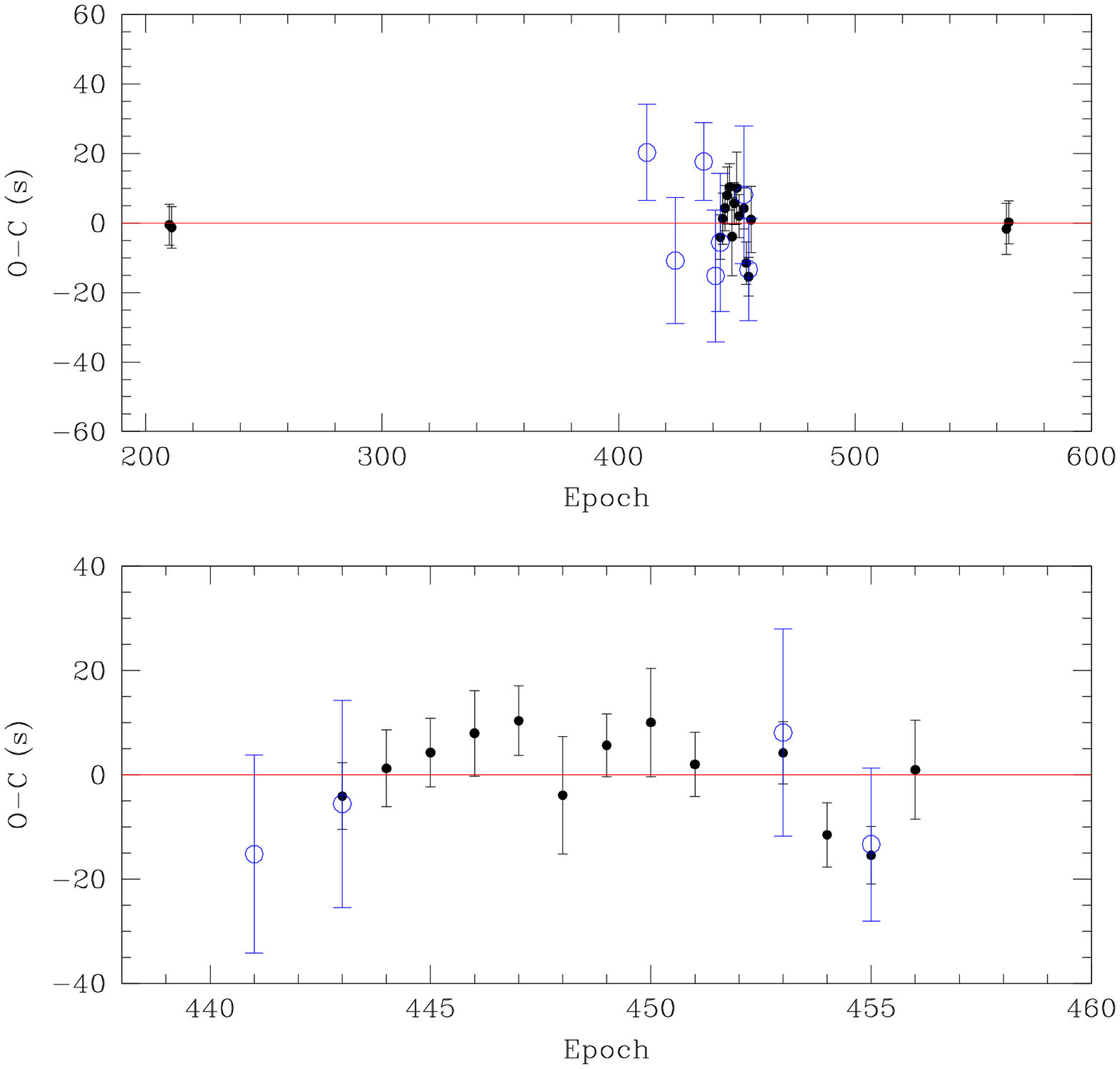}
\caption{$Top$: transit timing variations deduced from our global analysis for the {\it Spitzer} and TRAPPIST transits (see Sec.~3.4). 
$Bottom$: zoom on the consecutive transits observed by {\it Spitzer} in April and May 2011. Three of these transits were also observed by TRAPPIST.}
\end{figure}

\begin{table*}
\begin{center}
{\scriptsize
\label{tab:gj1214_3}
\begin{tabular}{cccc}
\hline\noalign {\smallskip}
& MCMC 1  & MCMC  2 & MCMC 3   \\ \noalign {\smallskip}
&  $e = 0$ & $e \ge 0$ & $e=0$, LD free  \\ \noalign {\smallskip}
\hline \noalign {\smallskip}
{\it Jump parameters} &  & \\ \noalign {\smallskip}
\hline \noalign {\smallskip}
$M_\ast$ [$M_\odot$]       & $0.176 \pm 0.009$ $(p)$ &  $0.176 \pm 0.009$ $(p)$ & $0.176 \pm 0.009$ $(p)$  \\ \noalign {\smallskip}
 $T_{eff}$ [K]                        & $3250  \pm 100$ $(p)$ &  $3250 \pm 100$ $(p)$ & $3250 \pm 100$ $(p)$                   \\ \noalign {\smallskip}
 $[Fe/H]$ [dex]                     &  $0.1 \pm 0.1$ $(p)$ &  $0.1 \pm 0.1$ $(p)$  &  $0.1 \pm 0.1$ $(p)$                  \\ \noalign {\smallskip}
$dF_{3.6\mu m}$ [\%]              &  $1.3545 \pm 0.0085$   &  $1.3521 \pm 0.0083$ & $1.342 \pm 0.015$\\ \noalign {\smallskip}
 $dF_{4.5\mu m}$ [\%]              &   $1.3676 \pm 0.0039$   &  $1.3673 \pm 0.0039$  & $1.365 \pm 0.011$     \\ \noalign {\smallskip}
$dF_{I+z}$ [\%]                          &  $1.377 \pm 0.020$  &     $1.376 \pm 0.021$  &$1.385 \pm 0.026$  \\ \noalign {\smallskip}
$dF_{occ,3.6\mu m}$ [ppm]    & $45_{-30}^{+44}$  & $74_{-51}^{+96}$    & $44_{-29}^{+44}$ \\ \noalign {\smallskip}
                                                      & $< 205$ (99.7\% confidence)  & $< 585$ (99.7\% confidence) &$< 184$ (99.7\% confidence)    \\ \noalign {\smallskip}
$dF_{occ,4.5\mu m}$ [ppm]     & $70 \pm 35$    &  $37_{-28}^{+41}$    & $67 \pm 35$        \\ \noalign {\smallskip}
                                                       & $< 190$ (99.7\% confidence)  &  $< 158$ (99.7\% confidence) & $< 177$ (99.7\% confidence)           \\ \noalign {\smallskip}
$A_{3.6\mu m}$ [ppm]               & $11_{-8}^{+15}$ & $16_{-12}^{+27}$   & $11_{-8}^{+16}$  \\ \noalign {\smallskip}
                                                       & $< 68$ (99.7\% confidence)  & $< 166$ (99.7\% confidence)  &  $< 78$ (99.7\% confidence)  \\ \noalign {\smallskip}
$O_{3.6\mu m}$ [deg]                & $350_{-110}^{+120}$  & $10_{-130}^{+100}$  &  $0_{-110}^{+120}$  \\ \noalign {\smallskip}
$A_{4.5\mu m}$ [ppm]                & $16_{-12}^{+18}$   & $11_{-8}^{+16}$ &  $18_{-13}^{+18}$ \\ \noalign {\smallskip}
                                                        & $< 68$ (99.7\% confidence)  & $< 68$ (99.7\% confidence)  &$< 73$ (99.7\% confidence)\\ \noalign {\smallskip}
$O_{4.5\mu m}$ [deg]                 & $325_{-110}^{+160}$                  	     & $335_{-110}^{+170}$ & $320_{-110}^{+180}$    \\ \noalign {\smallskip}
$b'$ [$R_\ast$]               &    $0.385 \pm 0.022$  &  $0.371_{-0.030}^{+0.017}$ &   $0.367 \pm 0.032$   \\ \noalign {\smallskip}
$W$ [min]                        &     $52.52 \pm 0.14$  &    $52.52 \pm 0.13$ &  $52.45 \pm 0.15$  \\ \noalign {\smallskip}
$T_0$  [BJD$_{TDB}$] & $2454980.74900 \pm 0.00010$  & $2454980.74901 \pm 0.00010$ & $2454980.74898 \pm 0.00008$\\ \noalign {\smallskip}
$P$ [d]                             & $1.58040417 \pm 0.00000022$  & $1.58040415 \pm 0.00000020$ &$1.58040421 \pm 0.00000018$  \\ \noalign {\smallskip}
$\sqrt{e}\cos\omega$   & 0 (fixed)  &  $-0.05 \pm 0.10$   & 0 (fixed)   \\ \noalign {\smallskip}
$\sqrt{e}\sin\omega$    & 0 (fixed)  &  $-0.13 \pm 0.21$   &   0 (fixed)  \\ \noalign {\smallskip}
$c_1(3.6{\mu}m)$         & $0.143 \pm 0.010$  &  $0.143 \pm 0.010$  & $0.204_{-0.087}^{+0.082}$    \\ \noalign {\smallskip}
$c_2(3.6{\mu}m)$         & $-0.392 \pm 0.010$  & $-0.391 \pm 0.010$ & $-0.77 \pm 0.54$  \\ \noalign {\smallskip}
$c_1(4.5{\mu}m)$         &  $0.189 \pm 0.010$  &  $0.189 \pm 0.010$    & $0.258 \pm 0.039$   \\ \noalign {\smallskip}
$c_2(4.5{\mu}m)$         &  $-0.3999 \pm 0.0070$ & $-0.3995 \pm 0.0070$ &   $-0.12_{-0.47}^{+0.34}$   \\ \noalign {\smallskip}
$c_1(I+z)$                      &  $0.749 \pm 0.063$ &  $0.753 \pm 0.068$   & $0.87 \pm 0.11$   \\ \noalign {\smallskip}
$c_2(I+z)$                      &  $-0.53 \pm 0.14$ &   $-0.54 \pm 0.15$   & $0.45_{-0.66}^{+0.63}$  \\ \noalign {\smallskip}
\hline \noalign {\smallskip}
{\it  Stellar parameters}  &   &  \\ \noalign {\smallskip}
\hline \noalign {\smallskip}
$R_\ast$ [$R_\odot$]            &  $0.2213 \pm 0.0043$ & $0.217_{-0.015}^{+0.010}$  & $0.2198 \pm 0.0045$\\ \noalign {\smallskip}
Luminosity $L_\ast$ [$R_\odot$]            &  $0.00488_{-0.00060}^{+0.00068}$ & $0.00465_{-0.00079}^{+0.00082}$ &  $0.00484_{-0.00059}^{+0.00064}$  \\ \noalign {\smallskip}
Density $\rho_\ast$ [$\rho_\odot$]    & $16.25 \pm 0.46 $   &  $17.1_{-1.6}^{+3.9} $ &  $16.53 \pm 0.57 $  \\ \noalign {\smallskip}
Surface gravity $\log{g}_\ast$ [cgs]                & $4.994 \pm 0.012$  & $5.010_{-0.035}^{+0.061}$ & $4.999 \pm 0.014$   \\ \noalign {\smallskip}
$u_1(3.6{\mu}m)$                 &  $-0.0210 \pm 0.0052$ $(p)$  & $-0.0209 \pm 0.0053$  $(p)$ & $-0.08_{-0.11}^{+0.14}$  \\ \noalign {\smallskip}
$u_2(3.6{\mu}m)$                 &  $0.1852 \pm 0.0050$ $(p)$  & $0.1852 \pm 0.0045$ $(p)$ & $0.34_{-0.20}^{+0.22}$ \\ \noalign {\smallskip}
$u_1(4.5{\mu}m)$                &  $-0.0046 \pm 0.0050$ $(p)$  & $-0.0044 \pm 0.0048$ $(p)$ & $0.08_{-0.10}^{+0.08}$  \\ \noalign {\smallskip}
$u_2(4.5{\mu}m)$                &  $0.1976 \pm 0.0030$ $(p)$ &  $0.1975 \pm 0.0029$ $(p)$ & $0.10_{-0.13}^{+0.19}$  \\ \noalign {\smallskip}
$u_1(I+z)$                             &  $0.193 \pm 0.049$ $(p)$ &  $0.194 \pm 0.049$ $(p)$ & $0.44_{-0.17}^{+0.16}$  \\ \noalign {\smallskip}
$u_2(I+z)$                             &  $0.363 \pm 0.063$ $(p)$ &  $0.366 \pm 0.060$ $(p)$  & $0 \pm 0.26$  \\ \noalign {\smallskip}
\hline \noalign {\smallskip}
{\it Planet parameters}  &  & \\ \noalign {\smallskip}
\hline \noalign {\smallskip}   
$(R_p/R_\ast)_{3.6\mu m}$       & $0.11638 \pm 0.00037$ & $0.11638 \pm 0.00035$&  $0.11587_{-0.00067}^{+0.00058}$  \\ \noalign {\smallskip}
$(R_p/R_\ast)_{4.5\mu m}$       & $0.11694 \pm 0.00017$   & $0.11693 \pm 0.00017$  & $0.11685_{-0.00052}^{+0.00043}$  \\ \noalign {\smallskip}
$(R_p/R_\ast)_{I+z}$                  & $0.11735 \pm 0.00086$   & $0.11732 \pm 0.00090$ & $0.1177 \pm 0.0011$   \\ \noalign {\smallskip}
$a/R_\ast$                                     & $14.45 \pm 0.15$  &  $14.7_{-0.5}^{+1.0}$ &$14.54_{-0.16}^{+0.18}$      \\ \noalign {\smallskip}
$a$ [au]                                         & $0.01488 \pm 0.00025 $   & $0.01489 \pm 0.00025 $  & $0.01486 \pm 0.00025 $\\ \noalign {\smallskip}
$i$ [deg]                                         & $88.47 \pm 0.10$                                    &  $88.56_{-0.16}^{+0.18}$  & $88.55 \pm 0.14$ \\ \noalign {\smallskip}
$e$                                                 & 0 (fixed) &  $0.054_{-0.044}^{+0.087} $   & 0 (fixed)   \\ \noalign {\smallskip}
$\omega$ [deg]                            & -     & $249_{-100}^{+47}$ & -   \\ \noalign {\smallskip}
$e\cos\omega$   & 0 (fixed)  &  $-0.007_{-0.023}^{+0.032}$   & 0 (fixed)   \\ \noalign {\smallskip}
$e\sin\omega$    & 0 (fixed)  &  $-0.026_{-0.065}^{+0.035}$   &   0 (fixed)  \\ \noalign {\smallskip}
$T_{eq}$ [K]$^a$                         & $604 \pm 19$   & $596_{-26}^{+24}$   & $603 \pm 19$\\ \noalign {\smallskip}
$R_{p, 3.6\mu m}$ [$R_\oplus$] & $2.805 \pm 0.056$    &$2.75_{-0.19}^{+0.11}$  &  $2.776 \pm 0.061$  \\ \noalign {\smallskip}
$R_{p, 4.5\mu m}$ [$R_\oplus$] & $2.821 \pm 0.056$    & $2.77_{-0.19}^{+0.11}$  &  $2.799 \pm 0.061$   \\ \noalign {\smallskip}
$R_{p, I+z}$ [$R_\oplus$]            & $2.830 \pm 0.062$    & $2.78_{-0.19}^{+0.12}$   & $2.823 \pm 0.069$  \\ \noalign {\smallskip}
\hline                          
\end{tabular}}
\end{center}
\caption{Median and 1-$\sigma$ limits of the marginalized a posteriori probability distributions
for the jump and system parameters from our global MCMC analysis of {\it Spitzer} and TRAPPIST
photometry (Sect. 3). The analysis we adopted is MCMC 1. $(p)$: a normal prior distribution was assumed 
(see text for details). $^a$Assuming a null Bond albedo 
and a homogeneous heat distribution between both hemispheres.  } 
\end{table*}

\begin{table}
\begin{center}
\label{tab:gj1214_4}
\begin{tabular}{cccc}
\hline\noalign {\smallskip}
Observatory & Channel & Epoch$^a$  & Mid-transit time  \\ \noalign {\smallskip}
                    &                 &                        &  ($BJD_{TDB}$)\\ \noalign {\smallskip}                    
\hline \noalign {\smallskip}
{\it Spitzer} &  $S1$         & 210     & $5312.633866 \pm 0.000069$ \\ \noalign {\smallskip}       
{\it Spitzer} &  $S2$          & 211    & $5314.214261 \pm 0.000069$ \\ \noalign {\smallskip}        
TRAPPIST &  $I+z$         & 412    & $5631.87575 \pm 0.00016$ \\ \noalign {\smallskip}              
TRAPPIST &  $I+z$        & 424     & $5650.84024 \pm 0.00021$ \\ \noalign {\smallskip}            
TRAPPIST &  $I+z$        & 436     & $5669.80542 \pm 0.00013$ \\ \noalign {\smallskip}              
TRAPPIST &  $I+z$        & 441     & $5677.70706 \pm 0.00022$ \\ \noalign {\smallskip}              
{\it Spitzer} &  $S2$         & 443   & $5680.867997 \pm 0.000074$ \\ \noalign {\smallskip}             
TRAPPIST &  $I+z$         & 443    & $5680.86798 \pm 0.00023$ \\ \noalign {\smallskip}             
{\it Spitzer} &  $S2$         & 444    & $5682.448463 \pm 0.000085$ \\ \noalign {\smallskip}              
{\it Spitzer} &  $S2$         & 445     & $5684.028902 \pm 0.000076$ \\ \noalign {\smallskip}              
{\it Spitzer} &  $S2$          & 446    & $5685.609349 \pm 0.000095$ \\ \noalign {\smallskip}              
{\it Spitzer} &  $S2$          & 447    & $5687.189781 \pm 0.000077$ \\ \noalign {\smallskip}              
{\it Spitzer} &  $S2$          & 448    & $5688.77002 \pm 0.00013$ \\ \noalign {\smallskip}              
{\it Spitzer} &  $S2$          & 449    & $5690.350535 \pm 0.000070$ \\ \noalign {\smallskip}              
{\it Spitzer} &  $S2$          & 450   & $5691.93099 \pm 0.00012$ \\ \noalign {\smallskip}              
{\it Spitzer} &  $S2$          & 451   & $5693.511301 \pm 0.000071$ \\ \noalign {\smallskip} 
{\it Spitzer} &  $S2$          & 453   & $5696.672135 \pm 0.000069$ \\ \noalign {\smallskip}  
TRAPPIST &  $I+z$          & 453   & $5696.67218 \pm 0.00023$ \\ \noalign {\smallskip}            
{\it Spitzer} &  $S2$          & 454   & $5698.252357 \pm 0.000071$ \\ \noalign {\smallskip} 
{\it Spitzer} &  $S2$          & 455   & $5699.832716 \pm 0.000064$ \\ \noalign {\smallskip}  
TRAPPIST &  $I+z$          & 455   & $5699.83274 \pm 0.00017$     \\ \noalign {\smallskip}  
{\it Spitzer} &  $S2$          & 456   & $5701.41331 \pm 0.00011$ \\ \noalign {\smallskip}       
{\it Spitzer} &  $S1$          & 564   & $5872.096930 \pm 0.000085$ \\ \noalign {\smallskip}   
{\it Spitzer} &  $S1$          & 565   & $5873.677356 \pm 0.000072$ \\ \noalign {\smallskip}          
\hline                             
\end{tabular}
\end{center}
\caption{Transit mid-times and 1$\sigma$ errors for the {\it Spitzer} and TRAPPIST 
transits from our global analysis (Sect. 3.4). $S1$ and $S2$ denote, respectively,
the 3.6 $\mu$m and 4.5 $\mu$m channels of {\it Spitzer}/IRAC. $^a$The epoch
is relative to the transit ephemeris shown in Table 3. } 
\end{table}

\section{Search for a second transiting planet}

We used the best-fit residuals {\it Spitzer} light curve obtained from our adopted global analysis to
perform a search for the transit(s) of a possible second planet. We did not use the TRAPPIST residuals
as their photometric precision is significantly weaker. Our residuals light curve contains 14,293 
photometric measurements. For each measurement, the error bar was multiplied by the
corresponding $\beta_w$ factor (see Table 1) to take into account the actual white noise budget
of the data. For each of the three channels, we also multiplied the error bars by the mean $\beta_r$
for this channel. We did not use for each light curve its derived $\beta_r$ shown in Table 1, as a larger $\beta_r$ 
could be due to a transit.

Our procedure was based on a search for periodic transit-like
signals over a grid of periods, phases, impact parameters, and depths. The probed periods
ranged from 0.1day  to 20.9 days, the period step being 0.0001days (8.6 s). For each period step, 
transit models centered at 100 evenly separated phases were compared to the period-folded 
light curve,  assuming a circular orbit, $M_\ast = 0.176$ $M_\odot$, and $R_\ast = 0.221$ $R_\odot$. For each phase,
 transit models with impact parameters of 0 and 0.5, and depths ranging from 100 ppm to 1000 ppm, were tested.
 For each period, the chi-square $\chi^2$ corresponding to the best-fitting transit profile in terms of phase, depth, and impact
 parameter was registered and compared to the chi-square assuming no transit.

Figure 6 presents the resulting transit periodogram. The strongest power peak corresponds to $P$=0.4157 days, 
 the improvement of the $\chi^2$ being 15.8. The corresponding folded light curve is
 also shown in Fig.~6.  With  14293 measurements and 4 more degrees of freedom
 for the transit model, a $\Delta \chi^2 = -15.8$ corresponds to a $\Delta BIC =  -15.8 + 4 \log(14293) = +22.5$.
 Using the BIC as a proxy for the model marginal likelihood, this $\Delta BIC$ results in a Bayes factor of 
 $e^{\Delta BIC/2} = 77000$ in favor of the no-transit model, translating into a false alarm probability (FAP) 
 of $\sim$99.999\%. This power peak in the periodogram has not yet represented a significant signal. 
 A simple computation shows that a $\chi^2$ improvement of $\sim$ -50 would be required
 to result in  a transit model one hundred times more likely than the no-transit model (FAP $\sim$ 1 \%). 
 
To better investigate the significance of the highest peak of our transit periodogram, we performed
 two MCMC global analyses similar  to the ones described in Sect. 3, each composed of two chains of 100,000 
 steps. The first MCMC  model included  only GJ\,1214\,b, assuming for it a circular orbit,  no TTV, no transit depth 
 chromaticity, no phase curve, and no occultation.  In the second MCMC, we added a second transit 
 planet in circular orbit with $P\sim0.4157$ day. From the two resulting BICs,  a Bayes factor
 was again computed. The advantage of this procedure is that it does not use the best-fit residual
  light curves for which a shallow transit signal could have been partially erased by the baselines detrending. 
 Furthermore, all the free parameters of the models have their a posteriori probability distributions probed in the same process, 
 ensuring a proper  error propagation. Compared to the model without a second transiting planet, the two-planet model is shown to 
 be $\sim$650 times less likely, confirming thus that the strongest peak in the periodogram shown in Fig. 6 does not 
 correspond to a significant transit signal. 
 
 To ensure that no periodic transit-like signal was missed by our  algorithm, we also analyzed our 
 {\it Spitzer}  residuals light curve with the BLS algorithm (Kov\'acs et al. 2002) available on the NASA Exoplanet Archive website\footnote{http://exoplanetarchive.ipac.caltech.edu}. Here, BLS found several power excesses at  short periods, 
 the most significant corresponding to 2.2025 days, its derived false alarm probability (FAP) being $\sim 1$ \%. 
 Once folded with this period, the residuals light curve shows a tiny transit-like signal
with an amplitude of $\sim$ 100 ppm (Fig.~7). Still, its duration is $\sim$ 5 hr, which is $\sim$ 5 times longer than expected
for the central transit of a planet in a 2.2-day circular orbit around GJ\,1214. This explains why our transit search algorithm
did not detect this possible signal. Nevertheless, we decided to 
better assess its reality by performing the same MCMC procedure
described above. Here too, the result is that the putative transit signal is not significant, the resulting 
Bayes factor being $10^{24}$ in favor of the model without a second transiting planet. 

We also noticed that several among the strongest BLS peaks corresponded  to flux {\it increases} 
instead of drops, suggesting that the forest of peaks at short periods is due to red noise of instrumental or 
astrophysical origin. From the injection of simulated periodic transits of different periods and depths in the raw photometry
and their analysis using the same procedure as described above (global modeling GJ\,1214\,b + systematics, 
transit search in the resulting residuals,  short MCMC  for the most significant detected signals), we concluded  
that only transits deeper than $\sim 200$ ppm 
(for periods $<$1 day) to $\sim 500$ ppm (for unique transits)  could be firmly detected in our {\it Spitzer} GJ\,1214 
data; this limitation comes from the combination of the photon noise ($\sim90$ ppm per hour) and the 
$50-100$ ppm  red noise present in the light curves. 
These limits correspond to a range in planetary radii of 0.35 - 0.5 $R_\oplus$.

For the sake of completeness, we also performed a visual search for  transit-like structures in the residual
light curves, but failed to detect anything convincing. We thus conclude that there is an absence of evidence 
for a second transiting planet around GJ~1214 in the {\it Warm Spitzer} photometry, while we would have clearly
detected any transit of a Mars-size or larger planet.

\begin{figure}
\label{fig:6}
\centering                     
\includegraphics[width=9cm]{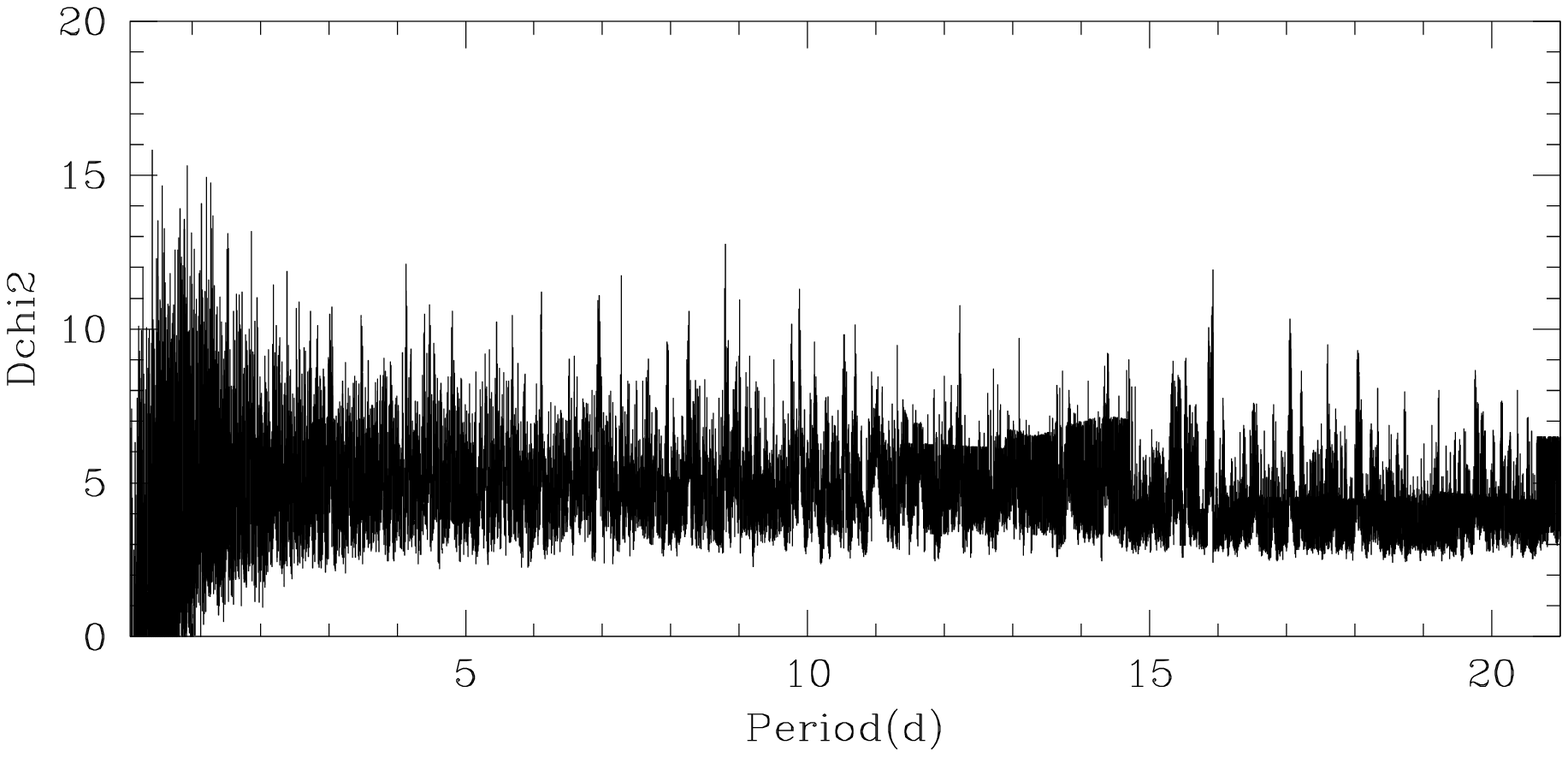}
\includegraphics[width=9cm]{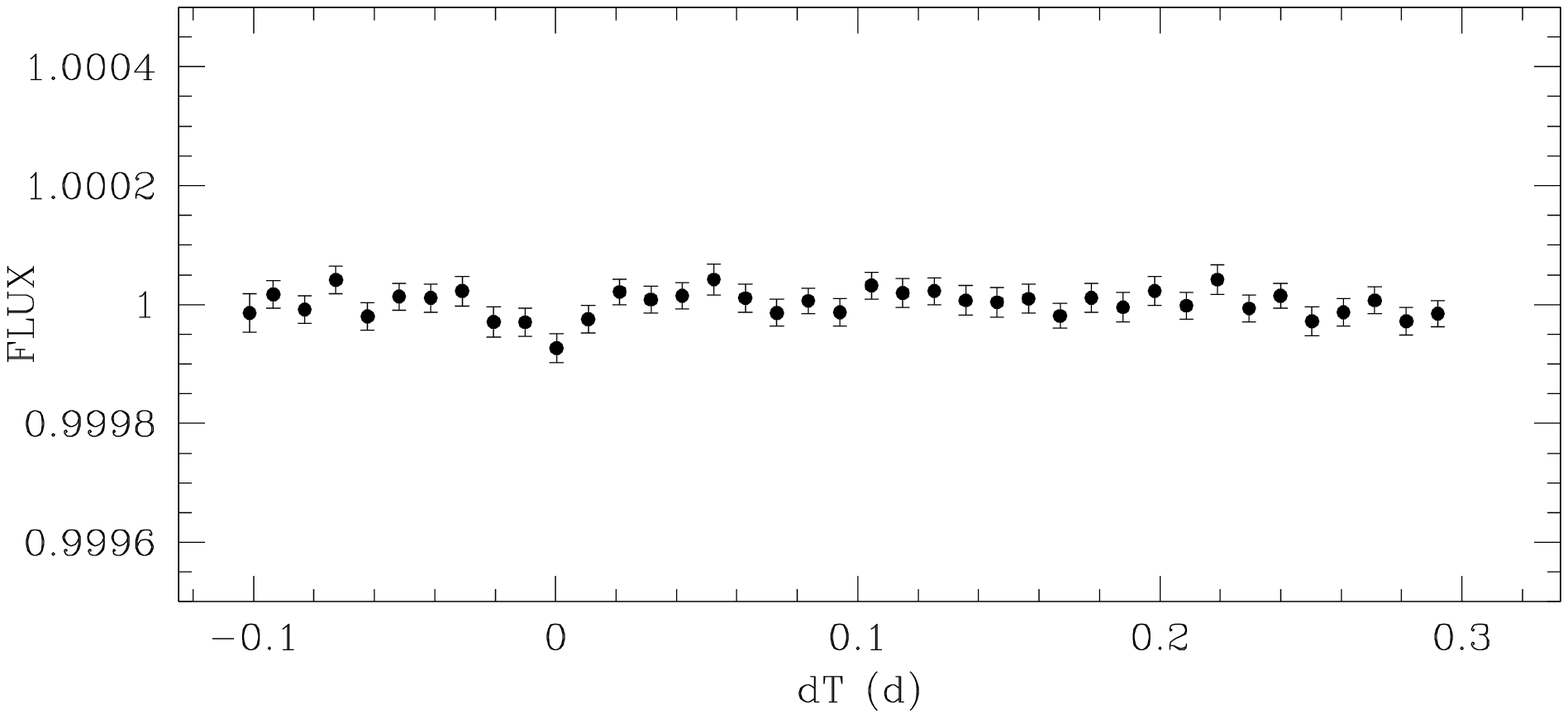}
\caption{$Top$: Transit search periodogram obtained from the analysis of the 
photometric residuals (Sect. 4). $Bottom$: {\it Spitzer} residuals folded on the ephemeris of the most significant
transit signal found by our transit search algorithm  ($P$=0.4157 day), and binned per intervals of 15 mins.}
\end{figure}

\begin{figure}
\label{fig:7}
\centering                     
\includegraphics[width=9cm]{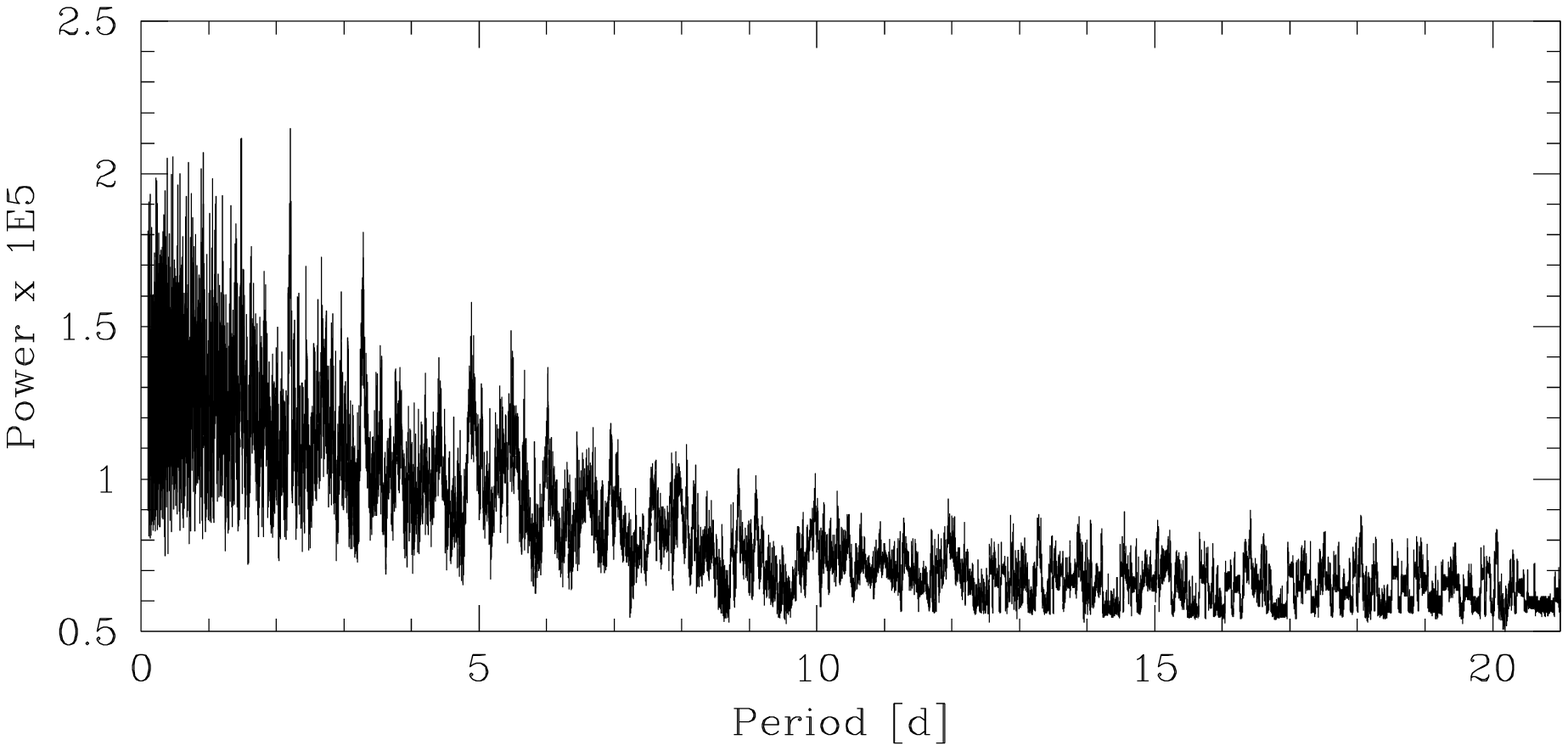}
\includegraphics[width=9cm]{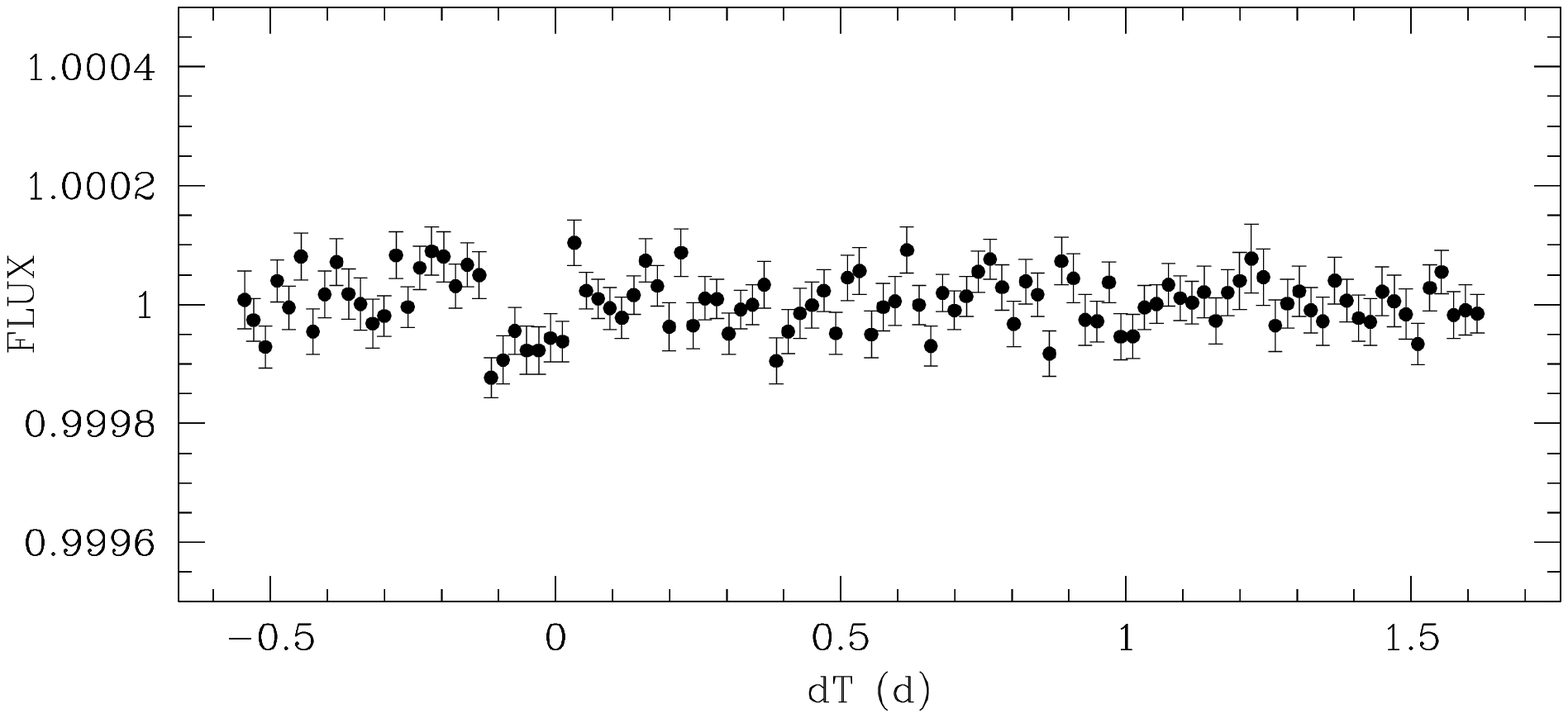}
\caption{$Top$: BLS periodogram obtained from the analysis of the {\it Spitzer}  
residuals light curve (Sect. 4). $Bottom$: {\it Spitzer} residuals folded on the ephemeris of the most significant
transit signal found by the BLS algorithm ($P$=2.2025 days), and binned per intervals of 30 mins.}
\end{figure}

\section{Discussion}

\subsection{The accuracy of our derived parameters}

The values deduced in our adopted analysis (MCMC 1 in Table 2) for the physical parameters $a/R_*$ 
and $i$  that define the transit shape (see, e.g.,  Winn 2011) are in good agreement with the values reported 
in the GJ\,1214\,b detection paper (C09), but disagree with several values reported afterwards and based
on high-precision photometry from the ground (Berta et al. 2011) or from space (Berta et al. 2012).  
Notably, the agreement with our own results presented in F13 is rather poor, while both analyses
were based on the same dataset.
This is illustrated in Fig.~8 ($left$). This figure shows a clear correlation between $a/R_*$ (and thus 
the stellar density) and the orbital inclination $i$. The simplest explanation to this correlation is that 
some of the shown measurements are affected by systematic errors. For a given orbital period, 
the same transit duration can be obtained from a larger star combined with a smaller inclination,  leading to a  
degeneracy between $a/R_*$ and $i$ that can be broken only by a very accurate determination of the transit shape. This is generally 
a difficult exercise because the red noise of the photometric time-series and the limb-darkening profile of the star 
alter the original trapezoidal profile of the transit.  Depending on the data quality and the details 
of the analysis, the derived values for $a/R_*$ and $i$ can easily be  affected by small systematic errors
 that are difficult to identify.  In F13, we  recognized the possibility of systematic errors in our
 analysis of the {\it Spitzer} data, so we forcibly varied $a/R_*$ to evaluate the effect on $R_p/R_*$ 
 (which was the most important parameter for that analysis).  These systematics in F13 could be due to the two-step approach 
 used in our analysis that started with the decorrelation of the raw 
photometry followed by the analysis of the detrended light curves. This strategy is not the best one to use for 
an accurate determination of transit parameters, as the initial decorrelation phase can slightly distort the 
transit shape and represents a source of error that is not explicitly propagated to the fitted transit parameters. 
The strategy used here that consists in the global modeling of the planetary and instrumental signals 
is better able to accurately determine the transit parameters. Nevertheless, we note that our main goal 
in F13 was  the accurate measurement of the transmission spectrum of the planet. This
goal was clearly achieved, as demonstrated not only by the agreement between  the transit depths 
measured in F13 for each channel with three different methods, but also by their excellent agreement with our
independent measurements presented  here (see Table 6).
 
Our multi-band global analysis strategy should be the best choice for breaking the  degeneracy between $a/R_*$ and $i$, 
notably by  averaging the influence of the instrumental systematics present in the different channels and by 
minimizing the impact of the limb-darkening uncertainties and the assumed model. Notably, this is supported 
by the results of our   MCMC 3 analysis that did not  assume any prior distribution on the limb-darkening coefficients and 
 still led to system parameters in excellent  agreement  ($< 1-\sigma$) with the ones derived in our adopted 
 MCMC 1 analysis  (see Table 2). To test further the reliability of our derived parameters,  for the three channels probed 
 by our data we modeled  the transits of a 2.8 $R_\oplus$ planet in front of a 0.176 $M_\odot$ - 0.221 $R_\odot$ star, assuming an orbital 
  period and inclination of $P$=1.58040417 d and $i=88.5^{\circ}$, respectively,  and quadratic limb-darkening coefficients 
  drawn from the Claret \& Bloemen (2011) tables for $T_{eff} = 3250$ K,    $\log{g} = 5.0$, and [Fe/H] = $0.1$. We injected the 
  corresponding transit profiles into our original light curves after dividing them by the best-fit transit models selected 
  by our MCMC 1 analysis. We then performed a new    MCMC analysis  similar to MCMC 1 in every way that resulted in parameter
   values fully consistent with the ones of MCMC 1.  Notably, we obtained $a/R_\ast = 14.52 \pm 0.15$ and $i=88.44 \pm 0.10$ deg,   
   in excellent agreement ($< 1 \sigma$)  with our input values. This test  suggests that our results for the transit shape parameters
    are accurate and do not suffer from strong systematic errors related to the details of our 
data analysis.
 
An actual chromatic variability of the transit shape could also explain the pattern visible in Fig.~8. It could be due to
 an inhomogeneous opacity of the planet limb at transit.   To test this hypothesis, we performed separate MCMC analyses of the 
  {\it Spitzer} 3.6 $\mu$m, 4.5 $\mu$m, and TRAPPIST $I+z$ photometry. The results are shown in Fig.~8 ($right$)
  and in Table 5. The parameters derived for the three channels are in excellent agreement with each other and with 
  the results of our global  MCMC 1 analysis. From this consistency, we conclude that systematic errors are the most
   plausible explanation for the  correlation between the measurements for $a/R_*$ and $i$ shown in Fig.~8 ($left$).

\begin{table}
\begin{center}
{\scriptsize
\label{tab:gj1214_3}
\begin{tabular}{cccc}
\hline\noalign {\smallskip}
 & $I+z$ & 3.6 $\mu$m  & 4.5 $\mu$m  \\ \noalign {\smallskip}
\hline \noalign {\smallskip}
$a/R_\ast$                                     & $14.64_{-0.67}^{+0.76}$     &  $14.62_{-0.33}^{+0.35}$     &$14.48_{-0.15}^{+0.17}$      \\ \noalign {\smallskip}
$i$ [deg]                                         & $88.76_{-0.52}^{+0.70} $    &  $88.57_{-0.23}^{+0.28}$    & $88.50 \pm 0.12$ \\ \noalign {\smallskip}
$R_p/R_\ast$                             & $0.1160 \pm 0.0017$           & $0.11629 \pm 0.00040$      & $0.11688 \pm 0.00018$   \\ \noalign {\smallskip}
\hline
\end{tabular}}
\end{center}
\caption{Median and 1-$\sigma$ limits of the marginalized a posteriori probability distributions for the parameters $a/R_\ast$, $i$, and $R_p/R_\ast$
derived from the separate analyses of the {\it Spitzer} 3.6 $\mu$m, 4.5 $\mu$m, and TRAPPIST $I+z$ photometry (Sect. 5.1).} 
\end{table}

\begin{table*}
\begin{center}
\label{tab:gj1214_3}
\begin{tabular}{ccccc}
\hline \noalign {\smallskip}
& This work & F13a & F13b & F13c  \\ \noalign {\smallskip}
\hline \noalign {\smallskip}
$(R_p/R_\ast)_{I+z}$                  & $0.11735 \pm 0.00086$                         & $0.1187 \pm 0.0011$                           & $0.1179 \pm 0.0018$ & $0.11803\pm0.00079$    \\ \noalign {\smallskip}
$(R_p/R_\ast)_{3.6\mu m}$       & $0.11638 \pm 0.00037$                         & $0.11607 \pm 0.00030$                      & $0.11616 \pm 0.00019$ & $0.11602 \pm 0.00055$     \\ \noalign {\smallskip}
$(R_p/R_\ast)_{4.5\mu m}$       & $0.11694 \pm 0.00017$                         &  $0.11710 \pm 0.00017$                    & $0.11699 \pm 0.00026$ & $0.11709 \pm 0.00022$     \\ \noalign {\smallskip}
\hline                          
\end{tabular}
\end{center}
\caption{Comparison of the planet-to-star radius ratio derived in this work and in F13 from the same data. F13a = simultaneous analysis for each channel; F13b = average of the results of individual analyses for each channel; F13c = analysis of the phase \& binned light curve for each channel. } 
\end{table*}

\begin{figure*}
\label{fig:8}
\centering                     
\includegraphics[width=8.5cm]{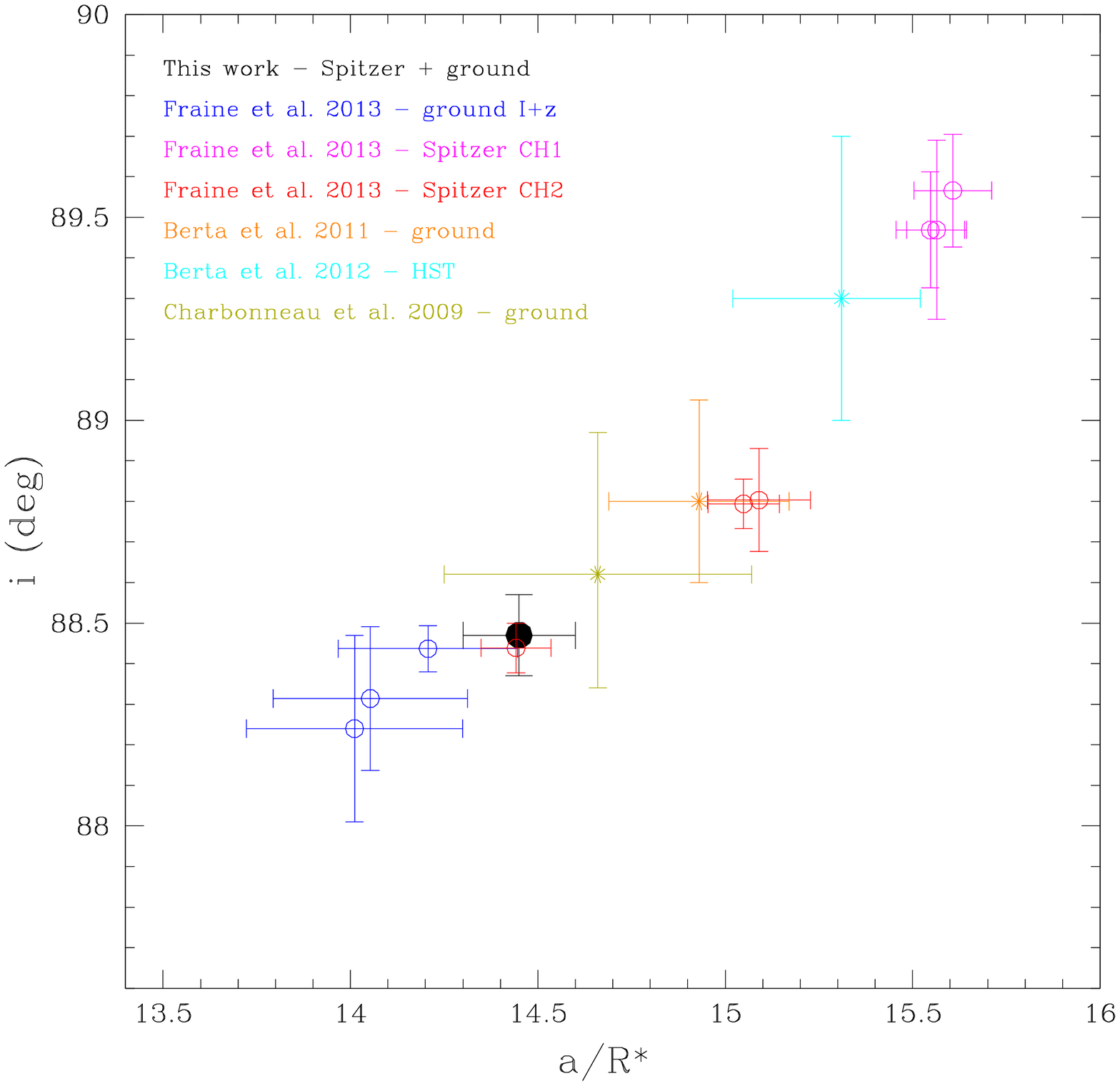}
\includegraphics[width=8.5cm]{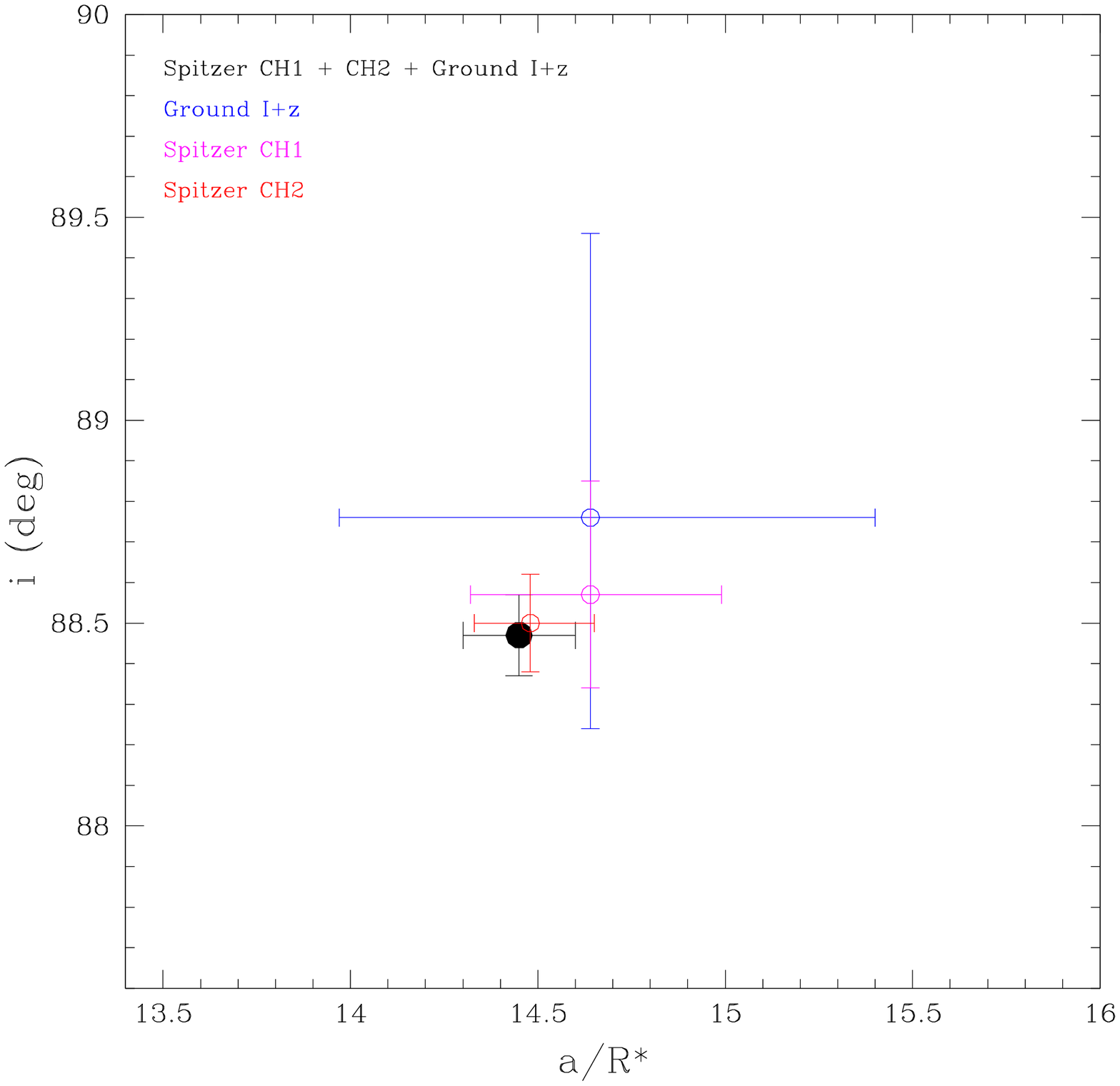}
\caption{Diagram of orbital inclination $i$ $vs.$ $a/R_\ast$ scale ratio. Our measurement is shown as a filled black circle. $Left$: comparison with the measurements presented by F13, C09 (ground), Berta et al. (2011) (ground), and Berta et al. (2012) (HST). $Right$: comparison of our measurement obtained from the global
analysis of {\it Spitzer} and TRAPPIST data to the measurements that we obtained from separate analyses of the {\it Spitzer}  3.6 $\mu$m, 4.5 $\mu$m, and 
TRAPPIST data alone. }
\end{figure*}

  \subsection{The presence of a habitable planet transiting GJ\,1214}
  
 The main motivation for our ambitious {\it Spitzer} program was to explore the  HZ of GJ\,1214 in search
  of a transiting planet, with a sensitivity high enough to detect any Mars-sized or larger planet. In practice, we did not probe
  the whole HZ, as  new estimates of its limits by Kopparapu et al. (2013) predict an outer edge of  0.13 au for GJ\,1214, 
  corresponding to a period of 44 days. In full generality, probing the entire HZ is in fact impossible, as the actual HZ outer limit 
 depends strongly on atmospheric composition (e.g., planets with hydrogen-dominated atmospheres could be habitable out
 to 1-2 au, Pierrehumbert \& Gaidos 2011), and possibly on the internal heat of the planets (Stevenson 1999). We thus  probed only the inner part of the 
 HZ, defined here as $P<20.9$ d. 
  
  We were able to reach the
  desired sensitivity, but unfortunately we did not detect a second planet transiting the red dwarf. Because we could not
  continuously monitor  GJ\,1214 during the 20.9 days, there is a small
  chance that we missed the transit of a second planet  orbiting in the inner part of the HZ, 
  especially if it is a planet with a longer period. To estimate this probability as a function
  of the orbital period, we  used the timings of the {\it Spitzer} data and determined for each orbital period the fraction of 
  orbital phases for which a mid-latitude transit would have happened at least once inside our observation window. 
  The resulting  probabilities are shown in Fig. 9.  For the inner part of the HZ, the mean probability to have observed
  at least one transit is  94\%, meaning that there is a 6\% chance that we missed the transit(s). So we cannot firmly
  reject (at 3-$\sigma$ or better)  that we missed the transit of a planet orbiting  in the inner part of the HZ, but the corresponding
  probability is certainly too low to justify additional monitoring of the system with a space telescope. For  orbits closer than  the HZ
  ($P<7.5$ d, using eq. 12 of Zsom et al. 2013 and our derived stellar luminosity $\sim0.0049$ $L_\odot$), the mean 
  probability to have observed at least one transit is 99.999\%, so we can firmly reject the presence of  a second Mars-sized 
  transiting planet for orbits closer than the HZ.
  
  Using a similar Monte-Carlo simulation procedure to the one described in G11, we derived the transit probability for a 
  putative second planet  taking into account the transiting nature of GJ\,1214\,b, using our new derived parameters for the stellar 
  size and for the orbital  inclination of GJ\,1214\,b, and drawing values out of the normal distribution $N(0,2.2^2)$ deg for
  the difference in inclination between both planets, 2.2 deg being the $rms$ of the inclinations of the solar system planets.
  This value is consistent with the spread in inclinations observed for {\it Kepler} multi-planetary systems (Fabrycky et al. 2012).
  The resulting probabilities are presented in Fig. 9 as a function of the orbital periods. The mean value for the inner part of the HZ
  is  27\%, and 61\% for the  zone closer than the HZ. Multiplying this geometric probability by the window probability derived above, and averaging  for the whole inner part of the HZ, we estimate that our {\it a priori} chance of success of detecting a habitable planet of Mars-size or above was 25\%, assuming GJ\,1214 does harbor such a planet  with $P<20.9$ d. Under this assumption, and taking into account our non-detection,  the  a posteriori probability that the planet does not transit is  $\sim 98$\%, while the  a posteriori probability that it does transit and that we missed  its transit is thus $\sim 2$\%. 
     
\subsection{The atmospheric properties of GJ\,1214\,b}

The atmosphere of GJ~1214\,b has been the subject of intense scrutiny in the recent past. Previous studies of the atmosphere of GJ~1214\,b have been based on observations of transmission spectra, which probe the regions near the day-night terminator of the planet. The sum total of existing data with multiple instruments over a wide spectral baseline ($\sim 0.8-5 \mu$m) indicate a flat transmission spectrum (Bean et al. 2010 \& 2011; D\'esert et al. 2011; Berta et al. 2011; de Mooij et al. 2012 \& 2013; Teske et al. 2013; but cf Croll et al. 2011). This spectrum is indicative of either a cloudy atmosphere with unknown composition (potentially H$_2$-rich) or an atmosphere with a high mean molecular weight ($\mu$), for example, an H$_2$O-rich atmosphere (Bean et al. 2010; D\'esert et al. 2011; Kempton et al. 2012; Benneke \& Seager 2012; Howe \& Burrows 2012; Morley et al. 2013). Constraining the atmospheric $\mu$ of GJ~1214\,b is important to address the fundamental question of whether super-Earths represent scaled-down Neptunes or scaled-up terrestrial planets. While a cloudy H$_2$-rich atmosphere would indicate a Neptune-like atmosphere for the planet, a high-$\mu$ atmosphere would suggest a terrestrial-like atmosphere. Current observations of the planetary atmosphere are unable to break the degeneracy between the two scenarios and so are inconclusive regarding the true composition of GJ~1214\,b.  

 Our observations of the secondary eclipses of GJ~1214b place constraints on the dayside atmosphere of the planet, a region not accessible to transmission observations. We use two photometric observations of the planet-star flux ratios in the 3.6 $\mu$m  and 4.5 $\mu$m bandpasses of {\it Spitzer}. We model the planetary thermal emission at secondary eclipse using the exoplanetary atmospheres modeling and retrieval method of Madhusudhan \& Seager (2009). The model computes line-by-line radiative transfer in a 1D plane-parallel atmosphere, with constraints of local thermodynamic equilibrium, hydrostatic equilibrium, and global energy balance. The pressure-temperature ($P-T$) profile and the molecular composition are free parameters of the model, allowing exploration of models with a wide range of temperature profiles (e.g., with and without temperature inversions) and chemical compositions (varied $\mu$, C/O ratios, etc.). However, given that we have only two photometric data points, compared to $\sim$10 free parameters depending on the specific model in question, the model space is presently under-constrained. We, therefore,  consider canonical models of the dayside atmosphere of GJ~1214\,b and investigate their potential in explaining the current data. We consider (1) H$_2$-rich solar composition models, (2) H$_2$O-rich models (called  water worlds), and (3) cloudy models parametrized by an optically thick cloud deck at a parametric pressure level. 

 Our results rule out a cloud-free solar abundance H$_2$-rich composition in the dayside atmosphere of GJ~1214\,b. In the temperature regime of GJ~1214\,b, as shown in the inset in Fig.~10, a solar abundance composition in chemical equilibrium predicts methane (CH$_4$) and water vapor (H$_2$O) to be the most dominant molecules bearing carbon and oxygen, respectively; CO$_2$ to be present at the $\sim$1 ppm level; and CO to be negligible (Madhusudhan \& Seager 2011; Madhusudhan 2012). The molecules CH$_4$ and H$_2$O have strong absorption in the 3.6 $\mu$m {\it Spitzer} channel, whereas CO and CO$_2$ have strong absorption features in the 4.5 $\mu$m {\it Spitzer} channel. Therefore, given their relative abundances, the corresponding model spectrum shows strong absorption in the 3.6 $\mu$m {\it Spitzer} channel and less absorption in the 4.5 $\mu$m channel, as shown in Fig~10. While this model spectrum explains the low planet-star flux contrast observed in the 3.6 $\mu$m, it predicts significantly higher contrast than is observed in the 4.5 $\mu$m channel. 

 On the other hand, our observations are consistent with both a metal-rich atmosphere and a cloudy H$_2$-rich atmosphere. Model atmospheres with a wide range of metal-rich compositions can explain the data. As shown in Fig~10, a water-world atmosphere (e.g., Miller-Ricci et al. 2009), with 99\% water vapor by volume fits both data within the 1-$\sigma$ uncertainties. The high mean-molecular weight of such an atmosphere causes a short atmospheric scale height, which together with the strong absorption features of water vapor cause low planet-star flux ratios across the near- to mid-infrared spectrum. This spectrum is consistent with the low planet-star flux ratios we observe in both the {\it Spitzer} channels at 3.6 $\mu$m and 4.5 $\mu$m. We also find that both the photometric data are consistent with a featureless blackbody spectrum of the planet with a temperature of $\sim$500-600 K, similar to a H$_2$-rich atmosphere with a gray-opacity cloud deck at pressures below $\sim$50 mbar. In this regard, our constraints on the composition of the dayside atmosphere of GJ~1214b are similar to the constraints on the atmospheric composition at the day-night terminator of the planet obtained from transmission spectra in the recent past (e.g., Bean et al. 2010, 2011; D\'esert et al. 2011; Berta et al. 2012).

\begin{figure}
\label{fig:9}
\centering                     
\includegraphics[width=8.5cm]{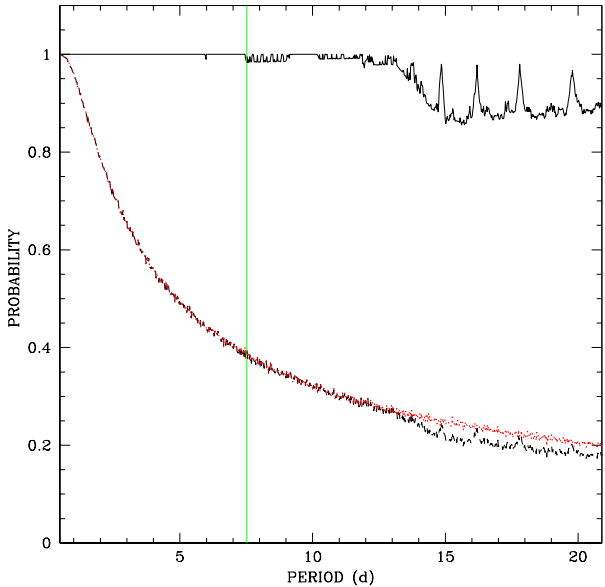}
\caption{Black solid line: probability for the transit of a second putative planet occurring at least once during the {\it Spitzer}  observations as 
a function of the orbital period, for periods up to  20.9 d. Black dots: transit probability for 
a second planet, taking into account the transiting nature of GJ\,1214\,b (see Sect. 5.2 for details). The red line shows the product of the 
two probabilities.  The green vertical line shows the inner limit of the HZ as computed from eq.~12 of Zsom et al. (2013).}
\end{figure}

\begin{figure}
\label{fig:9}
\centering                     
\includegraphics[width=8.5cm]{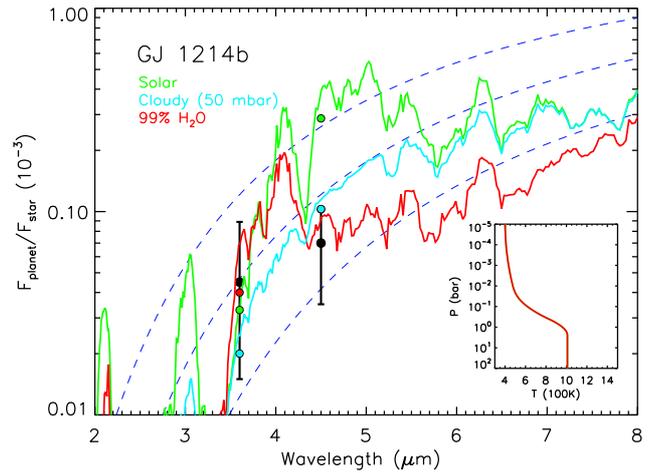}
\caption{Observations and model spectra of thermal emission from GJ~1214b. The black circles with error bars show the planet-star flux ratios observed in the {\it Spitzer} IRAC bandpasses at 3.6 $\mu$m and 4.5 $\mu$m. The green and red solid curves in the main panel show model spectra of an atmosphere with a solar abundance H$_2$-rich composition and one with a water-rich composition, respectively. The inset shows the temperature profiles for both models. The blue dashed curves show blackbody spectra of the planet with temperatures of 500 K, 600 K, and 700 K.}
\end{figure}

\section{Conclusions}

In G11, we had identified GJ\,1214 as a high-priority target for a transit search, as a habitable planet 
orbiting this nearby M4.5 dwarf should have its transit probability strongly enhanced by the transiting nature
of GJ\,1214\,b. In this context, we set up an ambitious high-precision photometric monitoring of GJ\,1214 
with the  {\it Spitzer Space Telescope} to probe  the inner part of its HZ ($P < 20.9$ d) in search of a planet as small as Mars. 
Because of a DNS failure, we could not probe the entire inner HZ, but still we covered about 94\% of it. 

After having presented in a first paper (F13) our detailed study of the {\it  Spitzer} transits of GJ\,1214\,b, and its
 implications for the transmission spectrum of the planet, we have reported here the results of our 
 transit search and of our global analysis of a very  extensive photometric dataset combining all the {\it Spitzer} 
 data acquired for GJ\,1214 to new ground-based transit light curves. Unfortunately, we did not detect a
 second transiting planet. Assuming that GJ\,1214 hosts a habitable planet larger than Mars  and with $P < 20.9$ d, our global analysis 
 of the whole {\it Spitzer} dataset leads to an a posteriori no-transit probability $\sim 98$\%.  Still, our analysis allowed 
 us to significantly improve the characterization of GJ\,1214\,b, notably by detecting at 2-$\sigma$ its 4.5 $\mu$m
  thermal emission, and by  constraining its 3.6 $\mu$m occultation depth to be smaller than 205ppm (3-$\sigma$ 
  upper limit). These emission measurements is new empirical evidence against a cloud-free hydrogen-rich 
  atmosphere for this intriguing super-Earth.  

 \begin{acknowledgements} 
This work is based in part on observations made with the {\it Spitzer Space Telescope}, which is operated by the Jet 
Propulsion Laboratory, California Institute of Technology, under a contract with NASA. This research has made use of the NASA
Exoplanet Archive, which is operated by the California Institute of Technology, under contract with the National
Aeronautics and Space Administration under the Exoplanet Exploration Program. M. Gillon and E. Jehin are Research Associates of
the Belgian Fonds National de la Recherche Scientifique (FNRS). L. Delrez is FRIA PhD student of the FNRS. 
A.~H.~M.~J. Triaud is a Swiss National Science Foundation fellow under grant number PBGEP2-145594.
TRAPPIST is a project funded by the FNRS under grant 
FRFC 2.5.594.09.F, with the participation of the Swiss National Science Fundation (SNF).  
The TRAPPIST team is grateful to Gregory Lambert and to the ESO La Silla  staff for their continuous support.  
N. Madhusudhan acknowledges support from Yale University through the YCAA postdoctoral prize fellowship. 
A. Zsom was supported by the German Science Foundation (DFG) under grant ZS107/2-1.  
\end{acknowledgements} 

\bibliographystyle{aa}

\end{document}